\documentclass[aps,showpacs,floatfix]{revtex4}
\usepackage[usenames]{color}
\usepackage{amssymb,epsfig}

\usepackage{graphicx}
\usepackage{bbm}
\usepackage{srcltx} 
\usepackage{subfigure}
\usepackage{amsmath,amsfonts}
\usepackage{slashed}
\usepackage{dcolumn}

\newcommand{\zsl}{\slashed{z}}

\newcommand{\ep}{\varepsilon}
\newcommand{\Dlr}{\stackrel{\leftrightarrow}{D}}
\newcommand{\bra}[1]{\left\langle #1 \right|}
\newcommand{\ket}[1]{\left| #1 \right\rangle}

\begin{document} 

\title{The Nucleon Distribution Amplitudes and their application to nucleon form
       factors and the $N \to \Delta$ transition at intermediate values of $Q^2$}

\author{Alexander Lenz, Meinulf G{\"o}ckeler, Thomas Kaltenbrunner, 
        Nikolaus Warkentin}

\affiliation{Fakult{\"a}t f{\"u}r Physik, Universit{\"a}t Regensburg, 93040 Regensburg, Germany}

\begin{abstract}
We compare a recent lattice determination of the nucleon distribution 
amplitudes with other approaches and models.
We study the nucleon distribution amplitudes up to twist 6 in next-to 
leading conformal spin and we also investigate conformal $d$-wave contributions
to the leading twist distribution amplitude.
With the help of light-cone sum rules one can relate the distribution amplitudes 
to the form factors of the nucleon or the $ N \to \Delta$ transition
at intermediate values of the momentum transfer. We compare our results with experimental
data in the range 1 GeV$^2 \leq Q^2 \leq 10$ GeV$^2$. 
Keeping in mind that we are working
only in LO QCD and NLO-QCD corrections might be sizeable we already  obtain a surprisingly 
good agreement for the nucleon form factors $G_M^n$, $G_M^p$, $G_A^p$ and $G_T^p$ 
and for the $N \to \Delta$  transition form factor ratios $R_{EM}$ and $R_{SM}$. 
\end{abstract}


\maketitle

\section{Introduction}
The nucleon distribution amplitudes represent the universal non-perturbative input
to numerous exclusive reactions, see, e.g., \cite{NDApioneer} for an early 
review.
Taking corrections up to twist-6 \cite{BFMS2000} into account we compare different 
non-perturbative methods to determine the nucleon distribution amplitudes,
in particular lattice simulations \cite{Lattice2007,Lattice2008,Lattice2008b}, QCD sum rule estimates 
\cite{BFMS2000,BLW2006} and a phenomenological model \cite{BLW2006}.
For asymptotically large values of the momentum transfer $Q^2$ the form factors can be 
expressed as a convolution of two leading-twist distribution amplitudes with a hard - 
perturbatively calculable - scattering kernel 
\cite{FFpQCD1,FFpQCD2,FFpQCD3,FFpQCD4,FFpQCD5,FFpQCD6,FFpQCD7,FFpQCD8,FFpQCD9}.
This approach (pQCD) is formally proven in the $Q^2 \to \infty$ limit, and currently there is
the consensus that pQCD is not valid at experimentally accessible values of the momentum transfer.
In \cite{BLMS2001} light-cone sum rules \cite{LCSR1,LCSR2} were worked out which  relate the
nucleon distribution amplitudes to the experimentally accessible form factors
of the nucleon at intermediate momentum transfer.
Form factors are interesting quantities per se, since they encode information
about the structure of the investigated baryon.
This interest rised a lot in recent years, in particular because new 
data from JLAB \cite{JLAB1,JLAB2,JLAB3,JLAB4} for the well-known 
electromagnetic form factors of the nucleon contradict common textbook-wisdom. 
See, e.g., \cite{FFreview} for a review and further references.
To our knowledge light-cone sum rules are the only theoretical approach to determine form
factors at intermediate momentum transfer  that incorporate consistently
the purely perturbative approach (pQCD). This was explicitly shown in the case of the pion form
factor \cite{pionFF1}. If one calculates the light-cone sum rules for the pion form factor
to leading order and next-to-leading order in QCD one can show that the $\alpha_s$-corrections
include the pQCD result in the $Q^2 \to \infty$ limit.
In the case of baryon form factors the pQCD result is expected to be included in the ${\cal O} (\alpha_s^2)$
corrections to the light-cone sum rule calculation.
Currently only leading-order sum rules for the baryon form factors are known and a part
of the NLO QCD corrections to the nucleon form factors.
\\
The paper is organized as follows.
In section 2 we introduce the concept of distribution amplitudes, in section 3 we collect 
QCD sum rule predictions for the nucleon distribution amplitudes and in section 4 we shortly explain the 
lattice determination of the moments of the nucleon distribution amplitudes.
All these approaches, including the numerical results, are discussed in section 5.
The light-cone sum rule formalism is introduced in section 6 where we also give 
a short overview over the current literature on light-cone sum rules for 
baryonic form factors. In the next three sections we compare light-cone sum rule predictions
with different models of the nucleon distribution amplitude for the form factors of the nucleon 
and for the $N \to \Delta$ transition. In section 7 we use the nucleon distribution amplitudes 
including next-to-leading conformal spin contributions to determine the
form factors, in section 8 we make use of some relations
between twist-4 and twist-3 parameters and in section 9 we investigate the effect
of the $d$-wave contributions to the leading twist distribution amplitude.
We conclude and summarize our results in section 11.
\\
In the appendix we give for the first time the full expression for all nucleon
distribution amplitudes up to twist-6 including also the $d$-wave contribution 
for the leading twist distribution amplitude.
\section{The nucleon distribution amplitudes}
The distribution amplitudes comprise the infrared behaviour in exclusive 
processes involving large momentum transfer. They remove the infrared 
divergences in the  perturbative diagrams encoding the nonperturbative
content of the process and are defined in terms of 
the Bethe-Salpeter wave function 
\begin{equation}
 \Psi_\textrm{BS}(x, k_\perp)=
\langle 0\vert T\left[q(x_1,k_{1,\bot})q(x_2,k_{2,\bot})q(x_3,k_{3,\bot})\right] \vert P
\rangle
\end{equation}
with $x_i$ being the longitudinal momentum fraction carried by the quark $i$, 
 $k_{i,\bot}$ its transverse momentum and $\vert P \rangle$ the 
nucleon state with momentum $P$ ($P^2 = M_N^2$). The distribution amplitudes are then 
obtained by integrating out the transverse momenta,
\begin{equation}
 \Phi(x_i,\mu)=Z(\mu)\int^{\vert k_\bot \vert\leq\mu} \mathrm d^2 k_{i,\bot}
\Psi_\textrm{BS}(x,k_\bot) \, ,
    \label{eq:darenscale}
\end{equation} 
where $Z$ results from the renormalisation of the quark field operators. 
In coordinate space the nucleon distribution amplitudes are derived from the 
following non-local nucleon-to-vacuum matrix element (here we follow the 
definitions  in \cite{BFMS2000})
\begin{equation}
\langle 0 |\epsilon_{ijk} 
u_{\alpha}^{i'} (a_1 x) [a_1 x, a_0 x]_{i',i}
u_{\beta }^{j'} (a_2 x) [a_2 x, a_0 x]_{j',j}
d_{\gamma}^{k'} (a_3 x) [a_3 x, a_0 x]_{k',k}
| P \rangle \, ;
\label{defDA}
\end{equation}
$u$ and $d $ are quark 
field operators, $\alpha$, $\beta$ and $\gamma$ are Dirac indices, while $i,
j$ and $k$ are color indices; $x$ is an arbitrary light-like vector,
$x^2 =0$, while the $a_i$ are real numbers that fulfill  $a_1+a_2+a_3=1$.
The gauge-factors $[x,y]$ are defined as
\begin{equation}
[x,y] = \mathcal P \exp 
\left[ 
i g \int \limits_0^1 dt (x-y)_\mu A^\mu (tx + (1-t)y)
\right] \, ,
\end{equation}
where path ordering $\mathcal P$ is implied.
They render the matrix element in Eq. (\ref{defDA}) gauge invariant.
In the following formulas we omit the gauge factors in order to simplify 
the notation.
\\
The leading-twist contribution to the nucleon distribution amplitudes has been
determined long time ago including terms of next-to-next-to leading conformal spin, 
see, e.g., \cite{NDApioneer} for an early review. We will compare different determinations
of the arising non-perturbative parameters in section 5.
Currently the nucleon distribution amplitudes have been expanded up to contributions
of twist 6 in \cite{BFMS2000} and the corresponding non-perturbative parameters were
estimated in \cite{BFMS2000, BLW2006} with QCD sum rules and in \cite{BLW2006} from a
phenomenological model. Some of these parameters were also calculated on the lattice 
\cite{Lattice2007,Lattice2008,Lattice2008b,Aoki06, Aoki08}. 
So-called $x^2$-corrections (corresponding to deviations from the light-like
separations of the quark fields in Eq. (\ref{defDA}))
to the leading twist distribution 
amplitudes were determined in \cite{BLMS2001,HW2004,BLW2006,L2007c}, they 
are formally of twist 5.
\\
Using the symmetry properties of the quark fields the matrix element in 
\eqref{defDA} can be expanded in twist as
\begin{equation}
4 \langle 0 |\epsilon_{ijk} 
u_{\alpha}^i (a_1 x)
u_{\beta}^j  (a_2 x)
d_{\gamma}^k (a_3 x) | P\rangle
= \sum \limits_i (\Gamma_3)^i_{\alpha \beta} (\Gamma_4)^i_{\gamma}\; F_i \, ,
\label{def1-NDA}
\end{equation}
where $\Gamma_{3/4}$ are certain Dirac structures and the $F_i$ are 
distribution amplitudes, which can be expanded into eigenstates of conformal
symmetry. This results in terms containing local operators. These local operators are associated 
with the moments of the distribution amplitudes, which are defined as
\begin{equation}
F^{n_1 n_2 n_3}_i = \frac{1}{F_{i,N} }\int \limits_0^1 \mathcal D x\; 
  x_1^{n_1}\, x_2^{n_2}\, x_3^{n_3}\; F_i (x_1, x_2, x_3).
\end{equation}
Here $F_i (x_1, x_2, x_3)$ stands for a distribution amplitude and $F_{i, N}$  for its 
normalisation constant, which is chosen such that $F_i^{000}\equiv 1$. The 
integration measure is defined as
\begin{equation}
 \mathcal D x= \mathrm d x_1\; \mathrm d x_2\; \mathrm d x_3\;
                         \delta(1-x_1-x_2-x_3).
\end{equation}  
Thus momentum conservation implies for the moments of the 
distribution amplitudes the relation
\begin{equation}
 F^{n_1n_2n_3}_i = F^{(n_1+1)n_2n_3}_i + F^{n_1(n_2+1)n_3}_i + F^{n_1n_2(n_3+1)}_i.
\end{equation}
Further details on distributions amplitudes (with complete expressions and definitions 
up to twist 6) are summarized in the appendices.
\\
For the  nucleon distribution amplitudes isospin symmetry and the 
presence of two quarks of the same type implies that the number of independent 
distribution amplitudes is reduced compared to the general case. In particular,
the leading-twist nucleon distribution amplitudes can be expressed in terms of
only one independent distribution amplitude which is usually taken as
\begin{equation}
 \varphi (x_1, x_2, x_3) = V_1 (x_1, x_2, x_3) - A_1 (x_1, x_2, x_3)
\end{equation}
and is equal to $\Phi_3(x_1, x_2, x_3)$ in the notation of \cite{BFMS2000}.
The distribution amplitudes $A_1$ and $V_1$ are defined in the appendices.
At leading twist the nucleon distribution amplitude $\varphi(x_i)$
corresponds to the following  form of the proton state
\cite{Chernyak1984,COZ1987}:
\begin{equation}
  \begin{split}
    \vert\, P,\uparrow\;\rangle=\int_0^1 \mathcal D x  \frac{\varphi(x_i)}{\sqrt{96 x_1 x_2 x_3}}
    &    
              \vert u^\uparrow(x_1) 
                     \left[
                            u^\downarrow(x_2) d^\uparrow(x_3) 
                          - d^\downarrow(x_2) u^\uparrow(x_3) 
                     \right]
                     \rangle\; .
  \end{split}
\end{equation}
The first moments of $\varphi(x_i)$ can be interpreted as the momentum fractions 
carried by the quarks. 
\\
The leading-twist distribution amplitude depends at leading conformal spin 
on one non-perturbative parameter, the normalization constant  $f_N$, while for 
twist four we have two additional constants  $\lambda_1$ and $\lambda_2$.
In our approach no new parameters appear in leading conformal spin up to twist six.
At next-to-leading conformal spin only two non-perturbative parameters 
$V_1^d=\varphi^{001}$ and $A_1^u=\varphi^{100}-\varphi^{010}$ arise in the 
case of leading twist and at next-to-leading twist we have three non-perturbative 
parameters, $f_1^d$, $f_1^u$ and $f_2^d$, for details see \cite{BFMS2000,BLW2006}.
For the leading-twist distribution amplitude $\varphi(x_i)$ we have also 
determined the next-to-next-to leading conformal spin contributions which can 
be completely parametrized, e.g., by the moments $\varphi^{101}$, $\varphi^{200}$ and 
$\varphi^{002}$. The local matrix elements defining the non-perturbative 
parameters up  to next-to leading conformal spin are (see \cite{BLW2006} for
the corrected formulas from \cite{BFMS2000})
\begin{align}
\langle 0 | \ep^{ijk} \left[u^i C\!\zsl\, u^j \right](0) \;
       [ \gamma_5 \zsl d^k]_{\delta}(0)  | P\rangle 
              & =
       { f_N} (P \! \cdot \! z) \zsl N_\delta (P)\label{eq:fn} \, ,
\\
\langle 0 | \ep^{ijk} \left[u^i C \gamma_\mu\, u^j\right](0) \;
       [\gamma_5 \gamma^\mu d^k ]_{\delta} (0)| P\rangle 
              & =
       { \lambda_1} m_N N_\delta (P) \, ,
\\
\langle 0 | \ep^{ijk} \left[u^i C \sigma_{\mu \nu}\, u^j \right](0) \;
       [\gamma_5 \sigma^{\mu\nu} d^k]_{\delta} (0) | P\rangle 
              & =
       { \lambda_2} m_N N_\delta (P) \, ,
\\
\langle 0 | \ep^{ijk} \left[u^i C\!\zsl\, u^j\right](0) \;
       [\gamma_5 \zsl  i z  \! \cdot \! \vec{D}d^k ]_{\delta}(0) | P\rangle 
              & =
       { f_N V_1^d} (P \! \cdot \! z)^2 \zsl N_\delta (P) \, ,
\\
\langle 0 | \ep^{ijk} \left[u^i C\!\zsl \, \gamma_5 i z \cdot \! \Dlr u^j \right] (0)\;
       [\zsl  d^k]_{\delta}(0) | P\rangle
              & =
       -{ f_N A_1^u} (P \! \cdot \! z)^2 \zsl N_\delta (P) \label{eq:A1u} \, ,
\\
\label{deff1d}
\bra{0} \ep^{ijk} \left[u^i C\gamma_\mu u^j \right](0)
       [\slashed{z} \gamma_5 \gamma^\mu  i z  \! \cdot \! \vec{D} d^k]_\delta(0) \ket{P}
              & =
       { \lambda_1 f_1^d} (P \! \cdot \! z) M \!\not\!{z} N_\delta (P) \, ,
 \\
\label{deff2d}
\bra{0} \ep^{ijk}   \left[u^i C\sigma_{\mu\nu} u^j \right](0)
       [\slashed{z} \gamma_5 \sigma^{\mu\nu} i z  \! \cdot \! \vec{D} d^k]_\delta (0) \ket{P} 
              & =
       { \lambda_2 f_2^d} (P \! \cdot \! z) M \slashed{z} N_\delta (P) \, ,
\\
\label{deff1u}
\bra{0} \ep^{ijk}  \left[u^i C \gamma_\mu \gamma_5 i z  \cdot \! \Dlr u^j\right](0)
       [\slashed{z}  \gamma^\mu d^k]_\delta(0) \ket{P}
              & =
       { \lambda_1 f_1^u} (P \! \cdot \! z) M \!\not\!{z} N_\delta (P) \, ,
\end{align}
with the nucleon spinor $N_\delta (P)$, the nucleon mass $m_N$, an arbitrary light-like vector
$z^\nu$ with $z^2 = 0$ and
$\Dlr= {\stackrel{\rightarrow}{D}} - {\stackrel{\leftarrow}{D}}$.
All derivatives act only on the quark fields and not on any explicit factor $z$.
The second moments of the nucleon distribution  amplitudes
are related to the following local operators: 
\begin{align}
\langle 0 | \ep^{ijk} \left[u^i C\!\zsl\, u^j\right] (0)\;
       [\gamma_5 \zsl\;  (i z  \! \cdot \! \vec{D})^2 d^k]_{\delta}(0) | P\rangle 
              & =
       { f_N \varphi^{002}} (P \! \cdot \! z)^3 \zsl N_\delta (P) \, ,
\\
\langle 0 | \ep^{ijk} \left[ ((i z  \! \cdot \! \vec{D})^2 u^i) C\!\zsl\, u^j \right] (0)\;
       [\gamma_5 \zsl d^k]_{\delta}(0) | P\rangle 
              & \nonumber \\
-
\langle 0 | \ep^{ijk} \left[ ((i z  \! \cdot \! \vec{D})^2 u^i) C\!\zsl \gamma_5  u^j\right](0) \;
       [\zsl  d^k]_{\delta}(0) | P\rangle
              & = 
       { f_N \varphi^{200}} (P \! \cdot \! z)^3 \zsl N_\delta (P) \, ,
\\
\langle 0 | \ep^{ijk} \left[ (i z  \! \cdot \! \vec{D} u^i) C\!\zsl u^j\right](0) \;
       [\gamma_5 \zsl i z \! \cdot \! \vec{D}  d^k]_{\delta}(0) | P\rangle 
              & \nonumber \\
-
\langle 0 | \ep^{ijk} \left[ (i z  \! \cdot \! \vec{D} u^i) C\!\zsl \gamma_5  u^j\right](0) \;
       [\zsl \; i z  \! \cdot \! \vec{D}  d^k]_{\delta}(0) | P\rangle
              & = 
       { f_N \varphi^{101}} (P \! \cdot \! z)^3 \zsl N_\delta (P) \, .
\label{eq:phi101}
\end{align}
The parameters used in this work with their twist and conformal spin are 
summarised in the table below
\begin{center}
\begin{tabular}{ | c|| c | c | }
\hline
                                   &   Leading twist                                   &    Higher twist \\ 
\hline
Leading conformal spin             &           $f_N$                                   &  $\lambda_1$, $\lambda_2$     \\
Next-to-leading conformal spin     &      $A_1^u$, $V_1^d$                             &  $f_1^u$, $f_1^d$, $f_2^d$    \\
Next-to-next-to-leading conformal spin & $\varphi^{101}$, $\varphi^{200}$, $\varphi^{002}$ &  --\\
\hline
\end{tabular}
\end{center}
As in the meson case these parameters can be estimated with QCD sum rules 
\cite{QCDSR} (see, e.g., \cite{BL2004,BBL2006,BBL2007} for some recent work in the meson case) 
or with lattice simulations (see, e.g., \cite{Lattice2006,Lattice2006b}
for lattice works considering the same mesons).

\section{QCD sum rule determination of the nucleon distribution amplitudes}
The leading-twist distribution amplitude was investigated with QCD sum rules up 
to the second moments in \cite{Chernyak1984,King1987} and up to the third moments
in \cite{COZ1987} including perturbative contributions and terms proportional 
to the gluon condensate and to the four-quark condensate (several errors 
in \cite{Chernyak1984} were corrected in \cite{COZ1987}).
\\
The next-to-leading twist normalization constants $\lambda_1$ and $\lambda_2$ 
describe the coupling to the proton of two independent proton interpolating 
fields used in QCD sum rules, $\lambda_1$ is the coupling of the so-called 
Ioffe current \cite{Ioffe1981}, while $\lambda_2$ is the coupling of the 
interpolating nucleon field that was advocated in \cite{Dosch1981}.
In \cite{Ioffe1981} and  \cite{Dosch1981} first QCD sum rule estimates for $\lambda_{1,2}$ 
were presented. Higher dimensional condensates were included in \cite{BelyaevIoffe1982}. 
Unfortunately these pioneering works contain several misprints, for a review with the correct 
expressions see, e.g., \cite{Leinweber1990,Ioffe2005}.
$\alpha_s$-corrections were calculated by Jamin in \cite{Jamin1988}. They turned out to be 
very large ($\approx + 50 \%$ for $|\lambda_1^2|$, 
corresponding to $\approx + 25 \%$ for $|\lambda_1|$), 
but we will not take them into account, since we also do not have 
$\alpha_s$-corrections for the light-cone sum rules, connecting the distribution amplitudes with 
the nucleon form factors. In \cite{BLW2006} also contributions of non-planar
diagrams to the dimension 8 condensates were included. Putting all this together (for the first time)
the QCD sum rule expression for $\lambda_1$ reads
\begin{eqnarray}
2 (2 \pi)^4 m_N^2 |\lambda_1^2| = & e^{\frac{m_N^2}{M_B^2}}& 
\left\{ M_B^6 E_3 \left( \frac{s_0}{M_B^2}\right) L^{-\frac49}
\left( 1 + \left[ \frac{53}{12} + \gamma_E \right] \frac{\alpha_s(M_B^2)}{\pi} \right)
\right.
\nonumber
\\
&+& \frac{b}{4} M_B^2  E_1 \left( \frac{s_0}{M_B^2}  \right) L^{-\frac49}
+ \left.
\frac{a^2}{3} \left[ 4 - \frac43 \frac{m_0^2}{M_B^2} \right]
\right\}
\label{lambda1} \, ,
\end{eqnarray}
where $M_B$ is the Borel parameter, $s_0$ is the continuum threshold and
\begin{eqnarray}
E_n(s_0/M^2) &= &1 - e^{(-s_0/M^2)} \sum_{k=0}^{n-1}
\frac{1}{k!} \left(\frac{s_0}{M^2}\right)^k \, ,
\\
L & = & \frac{\alpha_s(\mu^2)}{\alpha_s(M_B^2)} \, ,
\\
 a  & = &- (2\pi)^2 \langle \bar q q \rangle
\simeq 0.55 \; {\rm GeV^3} \,,
\\
b & = &(2\pi)^2 \langle \frac{\alpha_S}{\pi} G^2\rangle
\simeq 0.47 \; {\rm GeV^4} \,,
\\
m_0^2 & = & \frac{\langle \bar q g G q \rangle}{\langle \bar q
q\rangle}
\simeq 0.65 \; {\rm GeV^2} \,.
\end{eqnarray}
We have neglected in Eq.~(\ref{lambda1}) the small $\alpha_s$-corrections 
to the four-quark contribution proportional to $a^2$. The corresponding formula for $\lambda_2$ can be
found, e.g., in \cite{BLW2006}.
QCD sum rule estimates for the $f_x^y$ defined in Eqs.~(\ref{deff1d}) - (\ref{deff1u}) 
were first presented in \cite{BFMS2000} and updated in \cite{BLW2006}.
The parameter set which is obtained by QCD sum rules will be called {\it sum-rule} 
estimate in the following, we use the numerical values from \cite{COZ1987} for the moments
of the leading-twist distribution amplitude and the values from  \cite{BLW2006} for  
$f_N$, $\lambda_1$, $\lambda_2$, $A_1^u$, $V_1^d$, $f_1^d$, $f_1^u$ and $f_2^d$.
\\
In our analysis we use two related parameter sets which are based on the 
QCD sum rule determination:
\begin{itemize}
\item Demanding that all higher conformal contributions vanish, fixes 
      $A_1^u$, $V_1^d$, $f_1^d$, $f_1^u$ and $f_2^d$, while the values for
      $f_N$, $\lambda_1$, $\lambda_2$ are taken from the QCD sum rule estimates
      or from the lattice calculation.
      This parameter set will be called 
      {\it asymptotic}. In the case of the leading twist, one would be left with
      the asymptotic distribution amplitude  
      $\varphi(x_i,Q^2\rightarrow\infty)=\varphi_{asy}(x_i)=120 x_1 x_2 x_3 f_N$. 
       The corresponding expressions for the higher twist
      distribution amplitudes can be found in \cite{BFMS2000}.
\item With the help of light-cone sum rules \cite{BLMS2001,LWS2003,BLW2006}
      one can express the nucleon form factors in terms of the eight non-perturbative 
      parameters $f_N$, $\lambda_1$, $\lambda_2$, $A_1^u$, $V_1^d$, $f_1^d$, $f_1^u$ and $f_2^d$
      (inlcuding twist-6 corrections and expanding the distribution amplitudes up
       to the next-to-leading conformal spin). 
      Choosing values for these parameters in between the {\it asymptotic} and the {\it sum-rule} 
      values, 
      we got an astonishingly good agreement with the experimental numbers, see \cite{BLW2006}.
      This procedure is obviously rather ad-hoc and has to be replaced by a real fit
      after $\alpha_s$-corrections to the light-cone sum rules have been calculated.
      The paramter set obtained in \cite{BLW2006} will be called {\it BLW}.
\end{itemize}

\section{Lattice determination of the nucleon distribution amplitudes}

Lattice QCD offers the possibility to perform non-perturbative
computations in QCD without additional model assumptions. 
For example, one can evaluate hadron masses and matrix elements of 
local operators between hadron states. In particular, the 
non-perturbative parameters $f_N, \ldots$ introduced above can be 
extracted from Monte Carlo simulations on the lattice as advocated 
in \cite{Lattice1988}. 
\\
Recently, the QCDSF collaboration has performed such a calculation
\cite{Lattice2008}. It is based on gauge field configurations 
generated with two dynamical flavours of quarks. For the gauge field
the standard Wilson action was used, while the lattice action for the 
quarks was the so-called non-perturbatively $O(a)$ improved Wilson fermion 
action, also known as the clover fermion action. Although lattice artefacts
seem to be small, a reliable continuum extrapolation could not be 
attempted, and we utilize here the data obtained 
on the finest lattice corresponding to a gauge coupling parameter
$\beta = 5.40$. Setting the scale via a Sommer parameter of
$r_0 = 0.467 \, \mbox{fm}$ the lattice spacing turns out to be
$a \approx 0.067 \, \mbox{fm}$.
\\
On the lattice, matrix elements between the vacuum and a nucleon
such as those needed here are computed from two-point correlation
functions of the local operator $\mathcal O_\alpha (x)$ under study
and a suitable interpolating field $\bar{\mathcal N}_\alpha (x)$
for the nucleon. Asymptotically this two-point 
function decays exponentially with the distance between the operators since the
lattice calculations are performed in Euclidean space. Projecting
onto definite momentum one finds for sufficiently large (Euclidean)
times $t$:
\begin{equation}
\sum_{\vec{x}} \sum_{\vec{y}} e^{-i \vec{P} \cdot \vec{x}}
    e^{i \vec{P} \cdot \vec{y}}
    \langle \mathrm O_\alpha (\vec{x},t)
                          \bar{\mathcal N}_\beta (\vec{y},0) \rangle
= \frac{V_s \sqrt{Z}}{2 E(\vec{P}\,)} M_{\mathcal O}
  \left( E(\vec{P}\,) \gamma_4 - i \vec{P} \cdot \vec{\gamma} + m_N
         \right)_{\alpha \beta} e^{- E(\vec{P}\,) t} \,.
\end{equation}
Here $V_s$ denotes the spatial volume of the lattice and the matrix
elements of $\mathcal O_\alpha (x)$ and $\bar{\mathcal N}_\alpha (x)$
have been represented as
\begin{align}
\langle 0 | \mathcal O_\alpha (0) | P \rangle
                      = M_{\mathcal O} N_\alpha (P) \,, \\
\langle P | \bar{\mathcal N}_\alpha (0) | 0 \rangle
                      = \sqrt{Z} \bar{N}_\alpha (P) .
\end{align}
As the local operators $\mathcal O_\alpha (x)$ used 
in the simulations are linear combinations of the operators appearing in 
\eqref{eq:fn}-\eqref{eq:phi101}, the constants  $ M_{\mathcal O}$ are directly 
related to moments of the distribution amplitudes.
\\
The operators $\mathcal O_\alpha (x)$ need to be renormalized. In
Ref.~\cite{Lattice2008} a non-perturbative renormalization
procedure has been chosen. As the space-time symmetry on the lattice is reduced
to the finite (spinorial) hypercubic group, the mixing pattern
of our three-quark operators is more complicated than in the
continuum and the choice of the operators becomes an important
issue. Guided by the group-theoretical classification of three-quark
operators given in \cite{Kaltenbrunner:2008pb} 
the problematic mixing with lower-dimensional operators could however
be completely avoided. Moreover, the freedom in the choice of the 
operators has ben exploited in order to reduce the statistical 
uncertainties of the results. 
\\
Primarily, the combination of moments
\begin{equation}
 \phi^{n_1n_2n_3}=\frac{1}{3}\left(V_1^{n_1n_2n_3}-A_1^{n_1n_2n_3}
                                   +2\,T_1^{n_1n_3n_2}\right)
 =\frac{1}{3}\left(2\, \varphi^{n_1n_2n_3} + \varphi^{n_3n_2n_1}\right)
\end{equation}
has been evaluated, from which the combination $\varphi^{n_1n_2n_3}$ 
usually used in sum rule calculations is readily obtained by
\begin{equation}
 \varphi^{n_1n_2n_3}=2 \phi^{n_1n_2n_3}-\phi^{n_3n_2n_1}.
\end{equation} 
\\
In the following sections we shall compare these lattice results
with results obtained from other approaches and see what the lattice
numbers imply for the nucleon form factors.

\section{Comparison of different methods to determine the distribution amplitudes}
\begin{table}[ht]
 \begin{center}  
\renewcommand{\arraystretch}{1.25}
\fontsize{10.0pt}{12pt}\selectfont
\begin{tabular}{  | c|| D{.}{.}{3} || D{.}{.}{7} | D{.}{.}{3} | D{.}{.}{2} | D{.}{.}{3} | D{.}{.}{4} || D{.}{.}{13} |}
\hline
                    & \multicolumn{1}{c||}{Asy}& \multicolumn{1}{c|}{QCD-SR}& \multicolumn{1}{c|}{COZ} &  \multicolumn{1}{c|}{KS} & \multicolumn{1}{c|}{BK} & \multicolumn{1}{c||}{BLW} &  \multicolumn{1}{c|}{LAT}\\ \hline
$\varphi^{100}$ & \frac13\approx 0.333      & 0.560(60) & 0.579 & 0.55 & \frac{8}{21}\approx   0.38      & 0.415       & 0.3999(37)(139)    \\
$\varphi^{010}$ & \frac13\approx 0.333      & 0.192(12) & 0.192 & 0.21 & \frac{13}{42}\approx  0.31      & 0.285       & 0.2986(11)(52)    \\
$\varphi^{001}$ & \frac13\approx 0.333      & 0.229(29) & 0.229 & 0.24 & \frac{13}{42}\approx  0.31      & 0.300       & 0.3015(32)(106)    \\
\hline
$\varphi^{200}$ & \frac17\approx 0.143      & 0.350(70) & 0.369 & 0.35 & \frac{5}{28}\approx   0.18^\star& 0.225       & 0.1816(64)(212)    \\
$\varphi^{020}$ & \frac17\approx 0.143      & 0.084(19) & 0.068 & 0.09 & \frac{1}{8} \approx   0.13^\star& 0.121       & 0.1281(32)(106)    \\
$\varphi^{002}$ & \frac17\approx 0.143      & 0.109(19) & 0.089 & 0.12 & \frac{1}{8} \approx   0.13^\star& 0.132       & 0.1311(113)(382)   \\
$\varphi^{011}$ & \frac{2}{21}\approx 0.095 &-0.030(30) & 0.027 & 0.02 & \frac{1}{12}\approx   0.08^\star& 0.071       & 0.0613(89)(319)  \\
$\varphi^{101}$ & \frac{2}{21}\approx 0.095 & 0.102(12) & 0.113 & 0.10 & \frac{17}{168}\approx 0.10^\star& 0.097       & 0.1091(41)(152)    \\
$\varphi^{110}$ & \frac{2}{21}\approx 0.095 & 0.090(10) & 0.097 & 0.10 & \frac{8}{21}\approx   0.10^\star& 0.093       & 0.1092(67)(219)   \\
\hline
\end{tabular}
\end{center}
\caption{\label{tab:momcomp}
Comparison of different estimates for the moments of the leading-twist distribution amplitude 
renormalized at 1 GeV$^2$. 
We show the asymptotic values (Asy) and the QCD sum rule estimates from \cite{COZ1987}.
Inspired by the QCD sum rule calculation two models for the leading-twist distribution amplitude
were suggested, the COZ model \cite{COZ1987} and the KS model \cite{King1987}.
Using also some experimental input two phenomenological models were introduced, the BK model 
\cite{Bolz1996} and the BLW model \cite{BLW2006}. Finally we show the lattice values from
\cite{Lattice2007, Lattice2008, Lattice2008b}. The first error is statistical, the second error represents the uncertainty
due to the chiral extrapolation and renormalization.
For the BK model no contributions from 
next-to-next-to-leading conformal spin were  taken into account, thus the second moments denoted by the 
$^\star$ do not contain any additional information and are fully determined by the first moments.
}
\end{table}

In Table~\ref{tab:momcomp} we compare different estimates for the moments of the 
leading-twist distribution amplitude at 1 GeV$^2$. It turns out that the BLW model, the BK model 
and the lattice evaluation give almost indentical results, which are close to the asymptotic values,
while the QCD sum rule estimates seem to overestimate the deviation from the asymptotic form, although the deviation
goes in the right direction. The BLW model was inspired by this experience: One starts with the
asymptotic form and goes then in the direction of the QCD sum rule estimate, but only for a fraction
of the whole difference. Choosing this fraction to be 1/3 one gets an astonishingly good agreement
between light-cone sum rule predictions for the nucleon form factors and experiment, 
see \cite{BLW2006}.
In the same spirit one can make a BLW model for the second moments, also given in  
table~\ref{tab:momcomp} \footnote{To be precise, we have determined 
$\phi^{002}$, $\phi^{101}$ and $\phi^{110}$ 
from the BLW description and  $\phi^{200}$, $\phi^{020}$ and $\phi^{011}$ from momentum conservation.} .
These values are again very close to the lattice values.
The BK model \cite{Bolz1996} was also inspired by experiment, in particular
the decay $J / \Psi \to N \bar{N}$, 
the Feynman contribution to the nucleon form factor and the valence quark 
distribution function.
\begin{table}[ht]
\begin{center}
\fontsize{10.0pt}{12pt}\selectfont
 \begin{tabular}{  | c|| D{.}{.}{5} || D{.}{.}{6} | D{.}{.}{4} | D{.}{.}{5} || D{.}{.}{12} |}
\hline
                  & \multicolumn{1}{c||}{As.}& \multicolumn{1}{c|}{QCD-SR}& \multicolumn{1}{c|}{BK} &  \multicolumn{1}{c||}{BLW} &  \multicolumn{1}{c|}{LAT}\\ \hline
$ f_N\cdot 10^{3} [\mathrm{GeV}^2]$       &  5.0(5)                      &  5.0(5)                      & 6.64                       & 5.0(5)                       &  3.234(63)(86)\\
$\lambda_1\cdot 10^{3} [\mathrm{GeV}^2]$  &\multicolumn{1}{c||}{$-27(9)$}&\multicolumn{1}{c|}{$-27(9)$} &\multicolumn{1}{c|}{--}     &\multicolumn{1}{c||}{$-27(9)$}& -35.57(65)(136) \\
$ \lambda_2\cdot 10^{3} [\mathrm{GeV}^2]$ &\multicolumn{1}{c||}{$54(19)$}&\multicolumn{1}{c|}{$54(19)$} &\multicolumn{1}{c|}{--}     &\multicolumn{1}{c||}{$54(19)$}&  70.02(128)(268)\\
\hline
  $A_1^u $             & \multicolumn{1}{c||}{$0$}    &  0.38(15)                    &\frac{1}{14}\approx0.071 & 0.13            & 0.1013(81)(298) \\
  $V_1^d $             & \frac{1}{3}\approx0.333      &  0.23(3)                     &\frac{13}{42}\approx0.31 & 0.30                         & 0.3015(32)(106) \\
\hline
  $f_1^d$              & 0.30                         &  0.40(5)                     & \multicolumn{1}{c|}{--} & 0.33                         &\multicolumn{1}{c|}{--} \\
  $f_1^u$              & 0.10                         &  0.07(5)                     & \multicolumn{1}{c|}{--}    & 0.09                         &\multicolumn{1}{c|}{--} \\
  $f_2^d$              & \frac{4}{15}\approx 0.267    &  0.22(5)                     & \multicolumn{1}{c|}{--}    & 0.25                         &\multicolumn{1}{c|}{--} \\
\hline
\end{tabular}  
\end{center}
\caption{\label{tab:normcomp}
Comparison of different estimates for 
$f_N$, $\lambda_1$, $\lambda_2$, $A_1^u$, $V_1^d$, $f_1^d$, $f_1^u$ and $f_2^d$ renormalized at 
1 GeV$^2$. 
The QCD sum rule estimates and the BLW values are taken from \cite{BLW2006} and we  also show the 
phenomenological model from \cite{Bolz1996} (BK). For the asympotic and the BLW parameters 
the values for
$f_N$, $\lambda_1$ and $\lambda_2$ coincide with the ones from the QCD sum rule estimates.
}
\end{table}
In Table~\ref{tab:normcomp} we compare different estimates for 
$f_N$, $\lambda_1$, $\lambda_2$, $A_1^u$, $V_1^d$, $f_1^d$, $f_1^u$ and $f_2^d$ at $1 \, \mathrm{GeV}^2$. 
The leading twist parameters $A_1^u$, $V_1^d$, are already fully contained in  table \ref{tab:momcomp}
via the  relations
\begin{eqnarray}
A_1^u & = & 2 \varphi^{100} + \varphi^{001}-1 
\, ,
\\
V_1^d & = & \varphi^{001} 
\, .
\end{eqnarray}
Let us stress, however, that the errors quoted in Tables~\ref{tab:momcomp} and \ref{tab:normcomp} 
have to be taken with caution. 
On the lattice side a continuum extrapolation could not be attempted, and hence
the associated systematic error is not included.
Moreover, the errors on $A_1^u$ have been calculated by error propagation, which might not be too
reliable.
For the sum rule estimates the radiative corrections are expected to be sizeable,
but these are only known for $\lambda_1$ and $\lambda_2$.
The central lattice value of $f_N$ is about $35 \%$ smaller than the QCD  sum rule estimates, 
while the lattice results for $|\lambda_1|$ and  $|\lambda_2|$ 
are about $30 \%$  larger than the QCD  sum rule estimates.
For $\lambda_1$ and $\lambda_2$ the discrepancy is reduced strongly, 
if radiative corrections to the sum rule
estimates are included, cf. Eq.~(\ref{lambda1}), for $f_N$ - according to our knowledge - 
no radiative corrections have been calculated yet.
\\
The parameters $\lambda_1$ and $\lambda_2$ can also be extracted from the 
lattice calculation of the nucleon decay matrix elements (expressed in terms of the parameters
$\alpha$ and $\beta$) in \cite{Aoki06, Aoki08}.
Using the relations
\begin{equation}
\lambda_1 = \frac{4}{m_N} \alpha \, , \hspace{0.5cm}
\lambda_2 = \frac{8}{m_N} \beta \, ,
\end{equation}
we obtain from the results in \cite{Aoki06, Aoki08}
\begin{displaymath}   
\lambda_1 = -43.90 \pm 4.7 \pm 8.5 \cdot 10^{-3} \mbox{GeV}^2 \, , \qquad
\lambda_2 = 93.96  \pm 10.2\pm 22.7 \cdot 10^{-3} \mbox{GeV}^2 \, ,
\end{displaymath}
at the renormalization scale 1 GeV$^2$.
In that case the deviation from the QCD sum rule values is even more pronounced.
\\
In the non-relativistic  limit  one gets 
\begin{displaymath}
2 \lambda_1 + \lambda_2 = 0 \, .
\end{displaymath}
The estimates presented in table \ref{tab:normcomp} fulfill this relation
almost perfectly:
\begin{align}
\left \vert \frac{2 \lambda_1 + \lambda_2}{2 \lambda_1 - \lambda_2}\right \vert_{\mbox{QCD-SR}} & = 0 \pm 0.24\, ,
\\
\left \vert \frac{2 \lambda_1 + \lambda_2}{2 \lambda_1 - \lambda_2}\right \vert_{\mbox{LAT}} & =0.008 \pm 0.013\, .
\end{align}
For the ratio $f_N/\lambda_1$ the differences between the central lattice and QCD sum rule estimates
are even more enhanced:
\begin{eqnarray}
\left( \frac{f_N}{\lambda_1} \right)_{\mbox{QCD-SR}}  & = & -0.185 \pm 0.064 \, ,
\\
\left( \frac{f_N}{\lambda_1} \right)_{\mbox{LAT}} & = & -0.0909 \pm  0.0054 \pm  0.0095 \, .
\end{eqnarray}
The QCD sum rule estimate is a factor of two larger than the lattice result.
In the next section we will see that the electromagnetic form factors of the nucleon depend only on
the ratio but not on the individual values of $f_N$ and $\lambda_1$, if the so-called Ioffe 
interpolating field is used, while the $N \to \Delta$ transition depends on the individual values.
\section{Light-cone sum rules for form factors}
Light-cone sum rules (LCSR) are an advancement of QCD sum rules \cite{QCDSR} for intermediate
values of the momentum transfer $Q^2$, i.e., 1 GeV$^2 < Q^2 <$ 10 GeV$^2$ in the 
case of nucleon form factors. 
They were introduced in \cite{LCSR1,LCSR2}.
The starting point is a correlation function of the form
\begin{equation}
T (P,q) = \int d^4x e^{-iqx} 
\langle 0 | T \{ \eta(0) j(x) \} | N(P) \rangle \, ,
\label{corrstart}
\end{equation}
which describes the transition of a baryon $B$ with momentum $P-q$ to the nucleon $N(P)$
via the current $j$. The baryon $B$ is created  by the interpolating 
three-quark field $\eta$. If $B$ is a nucleon one can use, e.g., the Ioffe current 
\cite{Ioffe1981} for the proton
\begin{equation}
  \eta_{\rm Ioffe}(x) = 
\epsilon^{ijk} \left[u^i(x) (C \gamma_\nu)\, u^j(x)\right] \, 
(\gamma_5 \gamma^\nu)\, 
d^k_{\delta}(x) \, .
\label{Ioffecurrent}
\end{equation}
A typical example for $j$ is the electromagnetic current in the case of the
electromagnetic form factors
\begin{equation} 
j_{\mu}^{\rm em}(x) = e_u \bar{u}(x) \gamma_{\mu} u(x) + 
                       e_d \bar{d}(x) \gamma_{\mu} d(x) \, .
\label{emcurrent}
\end{equation}
With the definitions in Eqs.~(\ref{Ioffecurrent}),~(\ref{emcurrent}) the
correlation function in Eq.~(\ref{corrstart}) describes the electromagnetic 
form factors of the nucleon, which can be measured, e.g., in elastic
electron-proton scattering. 
\\
The basic idea of the light-cone sum rule approach is to calculate the
correlation function in Eq.~(\ref{corrstart}) both on the hadron level (expressed in terms of 
form factors) and on the quark level (expressed in terms of the nucleon
distribution amplitudes). Equating both results and performing a Borel 
transformation to suppress higher mass states one can express the
form factors in terms of the eight (taking only leading and next-to leading conformal spin into account)
non-perturbative parameters of the 
nucleon distribution amplitudes, the Borel parameter $M_B$ 
and the continuum threshold $s_0$, for details see \cite{BLMS2001,BLW2006}.
\\ 
We studied the electromagnetic nucleon form factors with the
Chernyak-Zhitnitsky interpolating field ($\eta_{CZ}$) in \cite{BLMS2001}.
In \cite{LWS2003} we found that $\eta_{CZ}$ yields  large unphysical isospin
violating effects, therefore we introduced a new isospin respecting CZ-like current to
determine the electromagnetic form factors.
In \cite{BLW2006} we also studied the Ioffe current for the nucleon and 
extended our studies from the electromagnetic form factors to axial
form factors, pseudoscalar form factors and the neutron to proton transition.
It turned out that the Ioffe current yields the most reliable results.
Despite our ``bad experience'' $\eta_{CZ}$ was used to determine the scalar 
form factor of the nucleon \cite{WW2006} and the axial and the pseudoscalar
one in \cite{WW2006b}.
The question of the ideal interpolating field can also be addressed more generally:
One can write down the most general interpolating field - without derivatives - 
of the nucleon as a linear combination
of two currents and then try to optimize the relative strength of these currents.
This approach was used for the scalar form  factor of the nucleon in \cite{AS2006},
for the axial-vector form factors in \cite{AS2007}
and for the electromagnetic form factors in \cite{Aliev2008}. Since in \cite{Aliev2008}
$x^2$-corrections were not included and different Dirac projections to extract the sum rules
were used, we cannot easily compare the result with \cite{BLW2006}.
\\
The light-cone sum rule method can also be applied to other observables than the nucleon 
form factors.
\\
In the class of nucleon to resonance transitions the following processes were considered:
The $N \to \Delta$ transition was studied in this framework in
\cite{BLPR2005} (for a similar approach for $Q^2 = 0$ see, e.g., \cite{R2007}),
the axial part of the  $N \to \Delta$ transition was calculated in
\cite{AAO2007}. Very recently the form factors of the $N \to N$*(1535) transition were
presented in \cite{Braun:2009jy}. In \cite{BILP2006, BIP2007, Braun:2009ya} pion-electroproduction 
was investigated.
\\
Also decays of baryons can be described with that formalism:
$\Lambda_b \to p l \nu$ was discussed in \cite{HW2004}. 
The authors of \cite{HW2006} considered  $\Lambda_c \to \Lambda l \nu$
and therefore determined a part of the $\Lambda$ distribution amplitude.
In \cite{W2006} the transition $\Sigma \to N$ was investigated. Recently
the rare decays $\Lambda_b \to \Lambda \gamma$ and  $\Lambda_b \to \Lambda l^+ l^-$ were 
treated in \cite{Wang2008} with the same formalism.
Electromagnetic form factors of $\Sigma$ and $\Lambda$-baryons were estimated in 
\cite{Liu:2008yg,Liu:2008zi}.
\\
So far all mentioned LCSR calculations for the baryon form factors were done in leading order
QCD. One expects sizeable radiative corrections of up to 30 $\%$. In \cite{Kornelija}
a first step in calculating the full ${\cal O} (\alpha_s)$-corrections to the nucleon
electromagnetic form factors was performed. The intrinsic final uncertainty of this 
approach is expected to be in the range of less than $\pm 20 \%$, 
if QCD corrections are included.
Comparing the theoretical predictions with experimental numbers 
one must be careful to distinguish  between quantities directly calculated like $F_1$ and $F_2$
and quantities like $G_E = F_1 - Q^2/(4 m_N^2) F_2$ for which cancellations might ruin the predictive power.
\\
In the following we use the LO QCD light-cone sum rules of \cite{BLW2006} for the electromagnetic form factors
of the nucleon and the LO QCD results of \cite{BLPR2005} for the $ N \to \Delta$ transition
to compare the consequences for the form factors which the lattice results for the nucleon distribution amplitudes 
entail with those which result from different QCD sum rule estimates.
Note, however, that the errors on the non-perturbative parameters of the nucleon distribution
amplitudes will not be taken into account, because this would not make much sense due to the
inherent uncertainty in the LO light-cone sum rules. 
%
%
%
%
%
%
%
%
%
%
%
%
\section{Results for the form factors at intermediate momentum transfer}
\begin{figure*}[ht]
\centering
       \subfigure{\includegraphics[width=0.315\textwidth,clip]{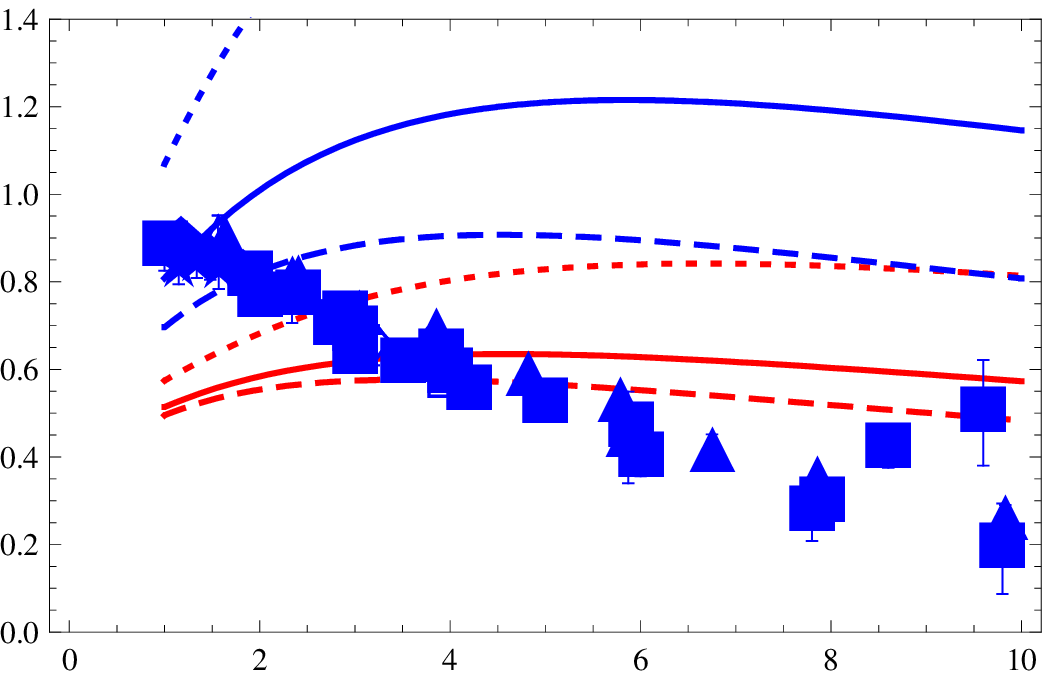}}
       \hspace{0.01\textwidth}
       \subfigure{\includegraphics[width=0.315\textwidth,clip]{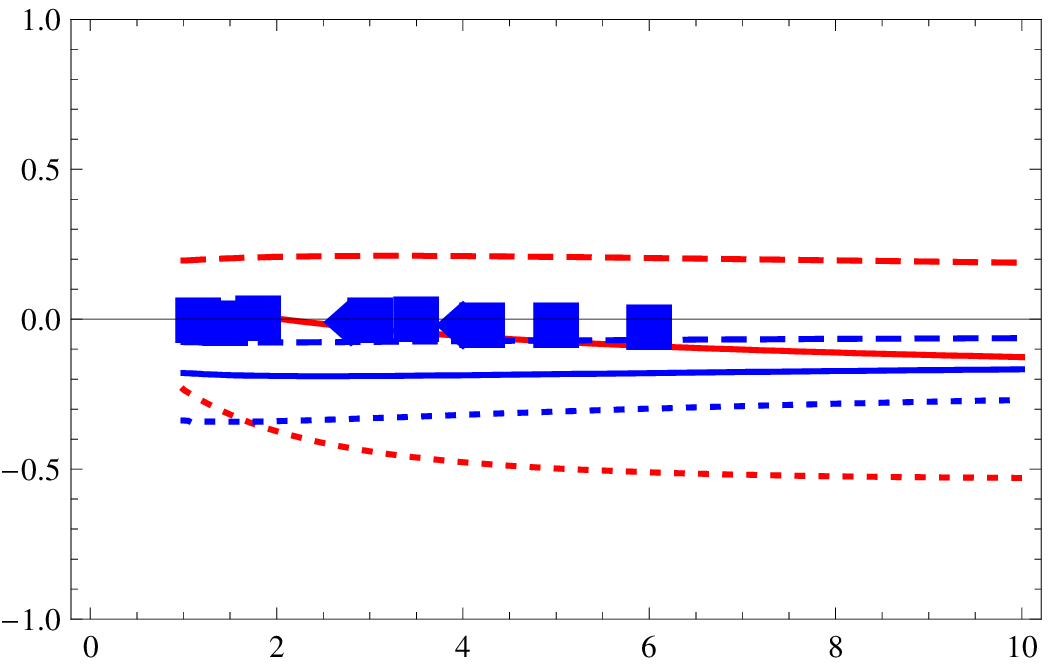}}
       \hspace{0.01\textwidth}
       \subfigure{\includegraphics[width=0.315\textwidth,clip]{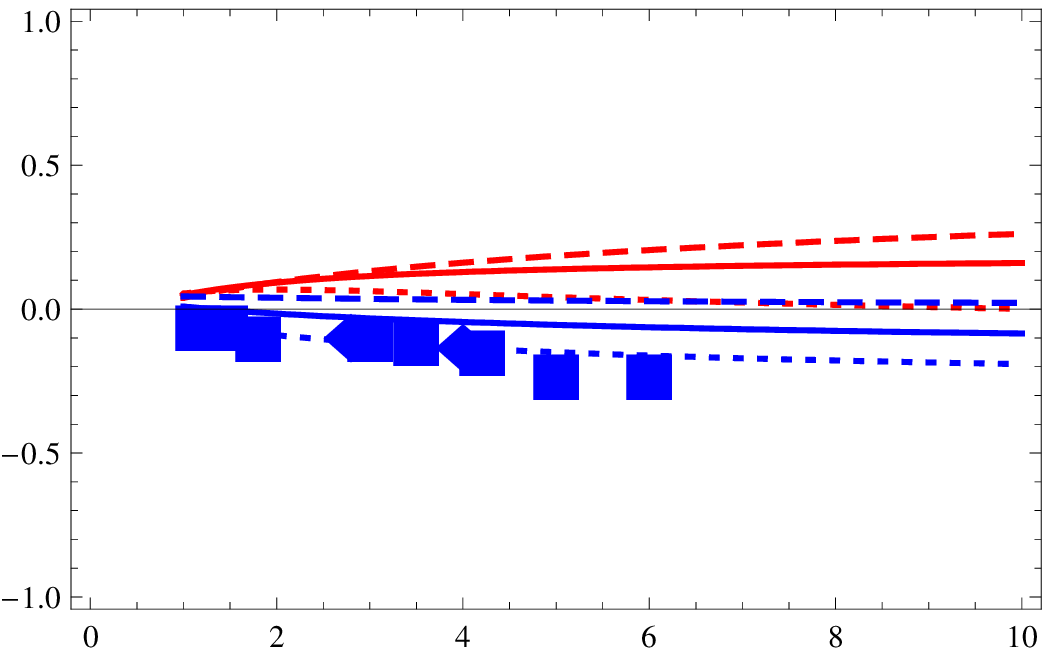}}
\caption{$\gamma^* N \to \Delta$ transition form factors 
(left:$G_M^*/(3 G_D)$ vs. $Q^2$, middle:$R_{EM}$  vs. $Q^2$, right: $R_{SM}$  vs. $Q^2$)
in  the LCSR approach~\cite{BLPR2005}
obtained using the BLW model (solid line), 
the asymptotic model (dashed line)
and the QCD sum rule model (dotted line) of the nucleon distribution amplitudes. 
The corresponding form factors based on the lattice values for the nucleon distribution 
amplitudes are given in blue. 
}
\label{fig:Delta}
\end{figure*}

In this section we use light-cone sum rules to extract physical form factors
from the nucleon distribution amplitudes, by taking into account conformal spin contributions
up to the p-wave; $d$-wave effects will be discussed in section 9.
\\ 
We compare our theory results to the following experimental numbers.
For the electromagnetic nucleon form factors we take data from:
\begin{itemize}
\item The magnetic form factor of the proton normalized to the dipole form factor
      $G_M^p/(\mu_p G_D)$ from \cite{ExpGMp1,ExpGMp2,ExpGMp3,ExpGMp4,ExpGMp5,ExpGMp6,ExpGMp7,ExpGMp8,ExpGMp9}, with
      \begin{equation}
      G_D(Q^2) = \frac{1}{\left( 1 + \frac{Q^2}{0.71 \mbox{GeV}^2} \right)^2}\, ,
      \hspace{1cm} \mu_p = 2.7928... 
      \end{equation}
      The data of \cite{ExpGMp3,ExpGMp4,ExpGMp5} are actually taken from the reanalysis in
      \cite{ExpGMp3-5real}.
\item The ratio of the electric and magnetic form factors of the proton $\mu_p G_E^p/G_M^p$ 
      from Rosenbluth separation 
      \cite{ExpGMp1,ExpGMp2,ExpGMp3,ExpGMp4,ExpGMp5,ExpGMp6,ExpGMp7,ExpGMp8,ExpGMp9,ExpGEGMp13,ExpGEGMp14,ExpGEGMp15} 
      and from polarization transfer \cite{JLAB1,JLAB2,JLAB3,JLAB4}. We would like to point out here that 
      \cite{ExpGMp8,ExpGMp9} claimed already in the seventies a steeper $Q^2$ dependence of $G_E^p$ compared to $G_M^p$ 
      for momentum transfers above 1 GeV$^2$. Currently the Rosenbluth separation data for $G_E$ are judged to be
      less reliable.
\item The ratio of the proton form factors $F_1^p$ and $F_2^p$ given as $\sqrt{Q^2} F_1^p/( (\mu_p-1) F_2^p) $ in  
      \cite{ExpF2F1p1,ExpGMp1,ExpGMp2}.
\item The magnetic form factor of the neutron normalized to the dipole form factor: $G_M^n/(\mu_n G_D)$ from 
      \cite{ExpGMn1,ExpGMn2,ExpGMn3,ExpGMn4,ExpGMp8,ExpGMn6,ExpGMn7,ExpGMn8} with $\mu_n = - 1.913...$ 
\item  The electric form factor of the neutron normalized to the dipole form factor: $G_E^n/       G_D $ 
      from  \cite{ExpGMn1,ExpGEn2,ExpGEn3,ExpGEn4,ExpGEn5,ExpGEn6,ExpGEn7,ExpGEn8,ExpGEn9,ExpGEn10,ExpGEn11,ExpGMn4,
                  ExpGEn13,ExpGEn14,ExpGEn15,ExpGEn16,ExpGMp8}. The data are very well described by the so-called Galster
      fit \cite{Galster}, we show in our plots the update of the Galster-fit from Kelly \cite{Kelly}:
      \begin{equation}
      G_E^{n,\it Galster}(Q^2) = 
     \frac{\left( 1.70 \pm 0.04\right) \tau }{1 + \left( 3.30 \pm 0.32\right) \tau} G_D(Q^2) \, ,
      \hspace{0.5cm} \mbox{with} \, \, \, \tau = \frac{Q^2}{4 m_p^2} \, .
      \end{equation}
\end{itemize}
For the axial form factors we compare our result to the dipole
formula \cite{ExpGAp1}
\begin{equation}
G_A(Q^2) = \frac{1.267}{\left(1+ \frac{Q^2}{(1.014 GeV)^2}\right)^2} \, .
\end{equation}
For more details see \cite{BLW2006}.
\\
Finally we use the following data for the $N \to \Delta$ transition:
\begin{itemize}
\item The magnetic form factor normalized to the dipole form factor $G_M^*/(3 G_D)$ from
      \cite{Stoler:1993yk,Stuart:1996zs,Bartel:1968tw,Stein:1975yy,Foster:1983kn,Kamalov:2000en,
            Alder:1972di,Frolov:1998pw}.
\item The ratio of the electric quadrupole to the magnetic form factor $R_{EM}$ from \cite{Joo:2001tw,Frolov:1998pw}.
\item The ratio of the Coulomb  quadrupole to the magnetic form factor $R_{SM}$ from \cite{Joo:2001tw,Frolov:1998pw}.
\end{itemize}
For more details see \cite{BLPR2005}.
Since we compare the data with the LCSR predictions, which are expected to work best in the region
1 GeV$^2 < Q^2 <$ 10 GeV$^2$, we are only interested in experiments where values of the form factors 
for momentum transfer above $Q^2 = 1$ GeV$^2$ are available.
\\
For the theory prediction we will use six models (i.e. six determinations for
the nonperturbative parameters
$f_N$, $\lambda_1$, $\lambda_2$, $A_1^u$, $V_1^d$, $f_1^d$, $f_1^u$ and $f_2^d$ ) 
for the nucleon distribution amplitudes
including s- and p-wave contributions:
\begin{enumerate}
\item QCD sum rule estimates (dotted red lines),
\item asymptotic form          (dashed red lines),
\item BLW model             (solid  red lines),
\item lattice evaluation plus QCD sum rule estimate for $f_x^y$ (dotted blue lines),
\item lattice evaluation plus asymptotic values     for $f_x^y$ (dashed blue lines),
\item lattice evaluation plus BLW          estimate for $f_x^y$ (solid  blue lines).
\end{enumerate}
Since $f_1^d$, $f_1^u$ and $f_2^d$ have not been determined on the lattice, we have to use in the 
lattice parameter set  QCD sum rule estimates, asymptotic values or the BLW model  for $f_x^y$.
For the nucleon form factors we use the LCSRs obtained in \cite{BLW2006} and for the $N \to \Delta$ transition 
we use the LCSRs obtained in \cite{BLPR2005}.
\begin{figure*}
       \centering
       \subfigure{\includegraphics[width=0.47\textwidth,clip]{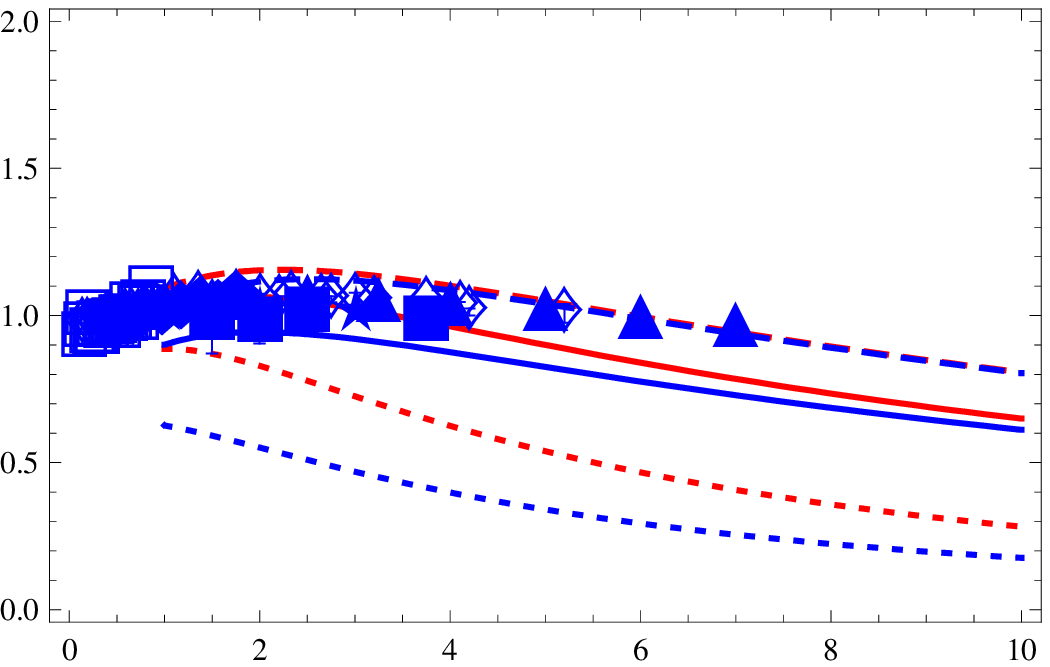}}
       \hspace{0.03\textwidth}
       \subfigure{\includegraphics[width=0.47\textwidth,clip]{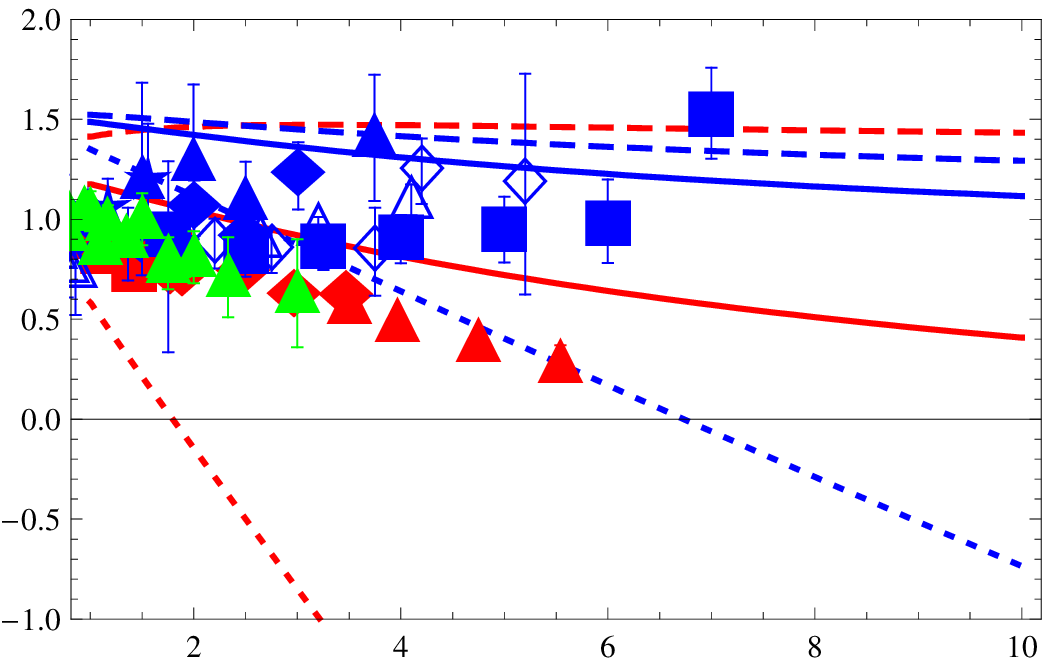}}
\caption{LCSR results for the electromagnetic form factors 
(left: $G_M/( \mu_p G_D)$ vs. $Q^2$; right:  $ \mu_p G_E/G_M$ vs. $Q^2$) of the 
proton, obtained using the BLW model (red solid line), 
the asymptotic model (red dashed line)
and the QCD sum rule model (red dotted line) of the nucleon distribution amplitudes. 
The corresponding results for the lattice values of the nucleon distribution amplitudes are given
in blue. The red data points on the right picture are JLAB data, while the blue and the green ones
are obtained via Rosenbluth separation. Currently the Rosenbluth separation data for $G_E$ are judged to be
less reliable.
}
\label{fig:GMGEp}
\end{figure*}
\begin{figure*}
\centering
\subfigure{\includegraphics[width=0.47\textwidth,clip]{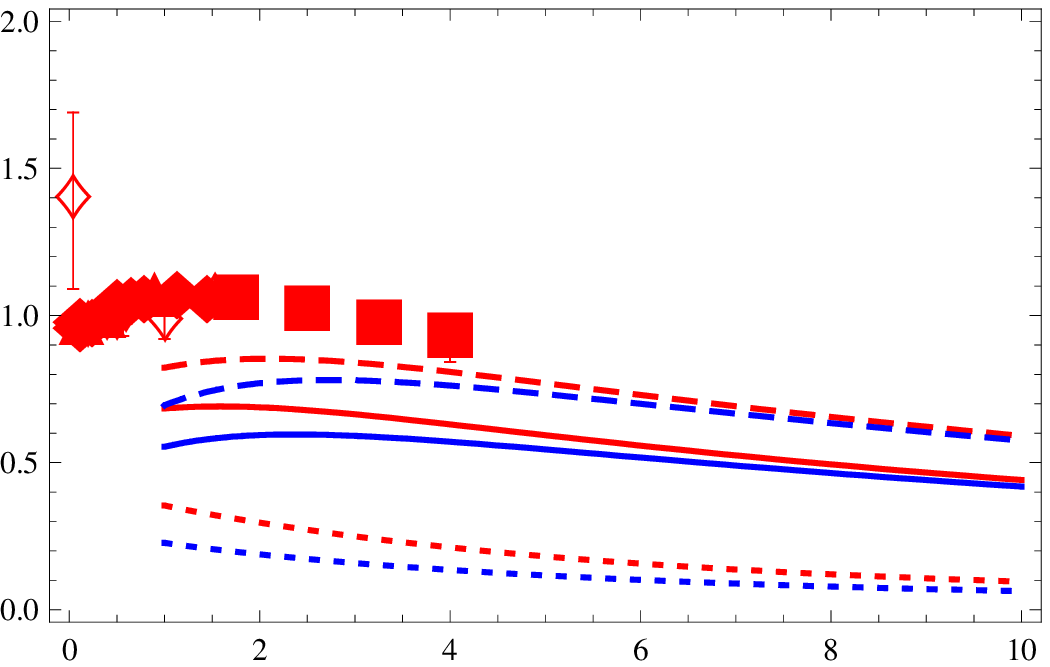}}
\hspace{0.03\textwidth}
\subfigure{\includegraphics[width=0.47\textwidth,clip]{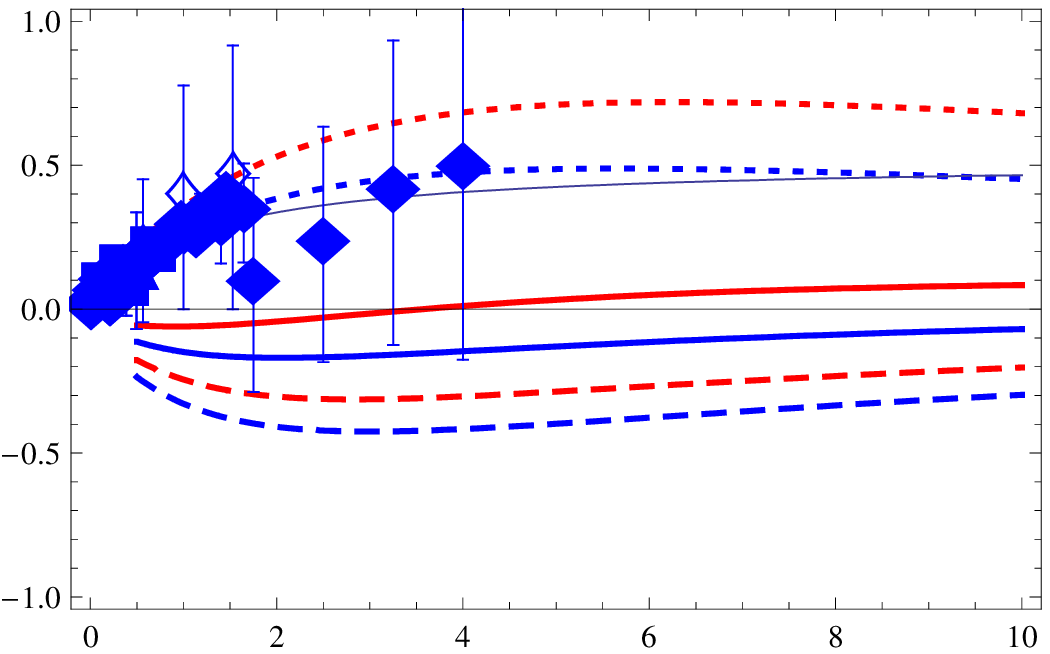}}
\caption{LCSR results for the electromagnetic form factors of the 
neutron (left: $G_M/(\mu_n G_D)$ vs. $Q^2$; right:  $G_E/(G_D)$ vs. $Q^2$), 
obtained using the BLW model (red solid line), 
the asymptotic model (red dashed line)
and the QCD sum rule model (red dotted line) of the nucleon distribution amplitudes. 
The corresponding results for the lattice values of the nucleon distribution amplitudes are given
in blue. The thin solid blue line represents the updated Galster fit.
}
\label{fig:GMGEn}
\end{figure*}
\begin{figure*}
       \centering
       \subfigure{\includegraphics[width=0.47\textwidth,clip]{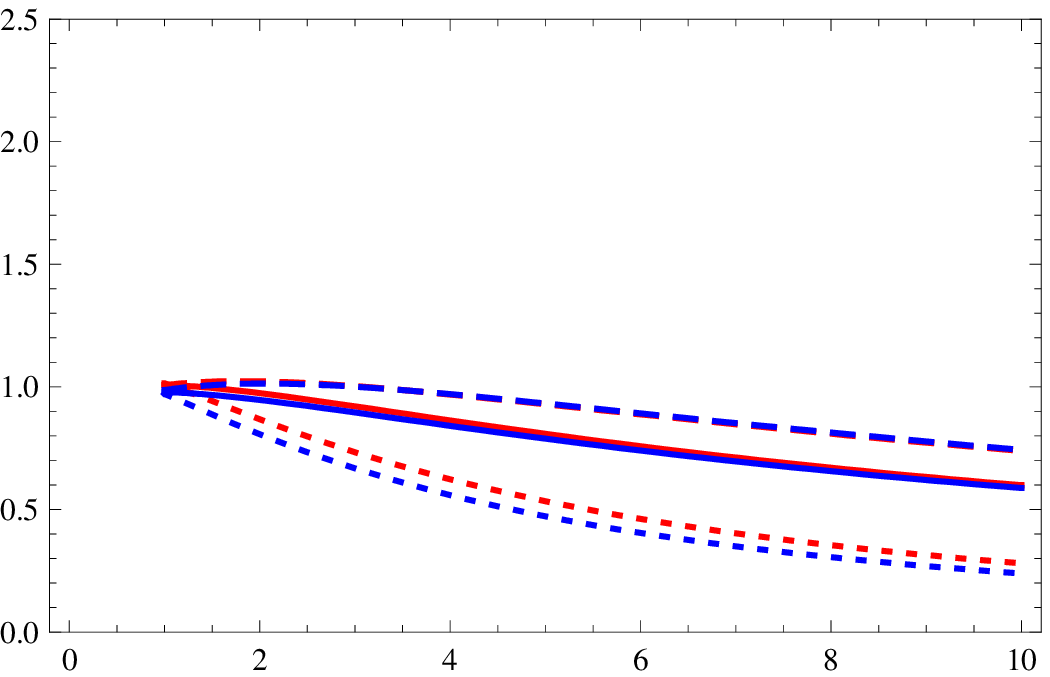}}
       \hspace{0.03\textwidth}
       \subfigure{\includegraphics[width=0.47\textwidth,clip]{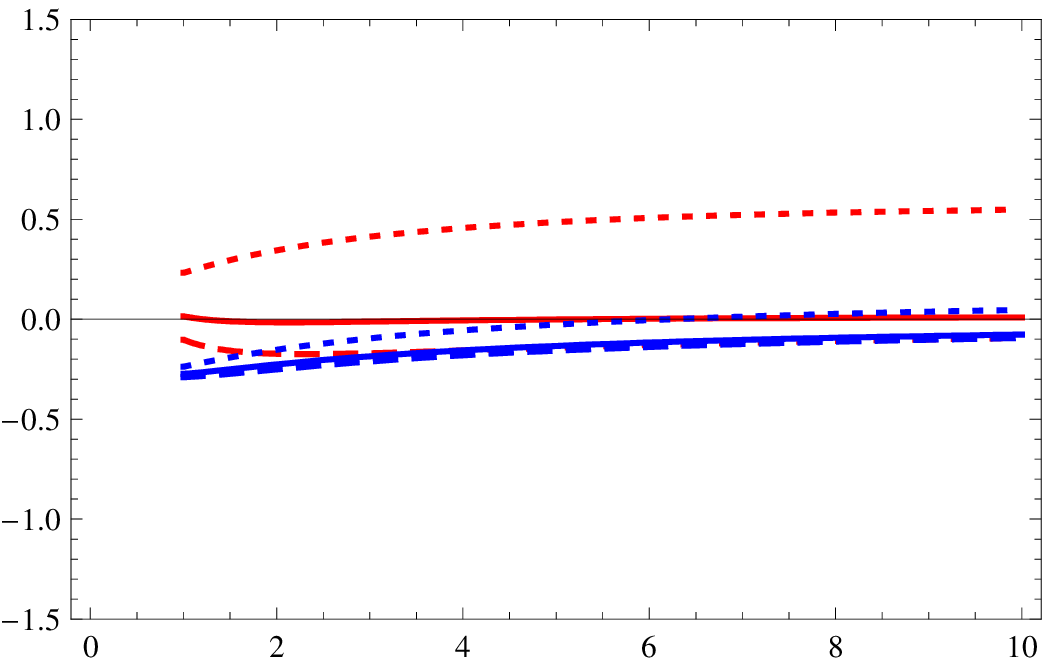}}
\caption{LCSR results (solid curves) for the axial form factor  of the proton
$G_A$ normalized to $G_D=g_A/(1+Q^2/m_A^2)^2$  vs. $Q^2$(left panel) and 
the tensor form factor $G_T$ normalized to $G_A$  vs. $Q^2$ (right panel),
obtained using the BLW model (red solid line), 
the asymptotic model (red dashed line)
and the QCD sum rule model (red dotted line) of the nucleon distribution amplitudes. 
The corresponding results for the lattice values of the nucleon distribution amplitudes are given
in blue. 
}
\label{fig:GAGT}
\end{figure*} 
\begin{figure}[ht]
\centering
  \includegraphics[width=0.8\textwidth,angle=0]{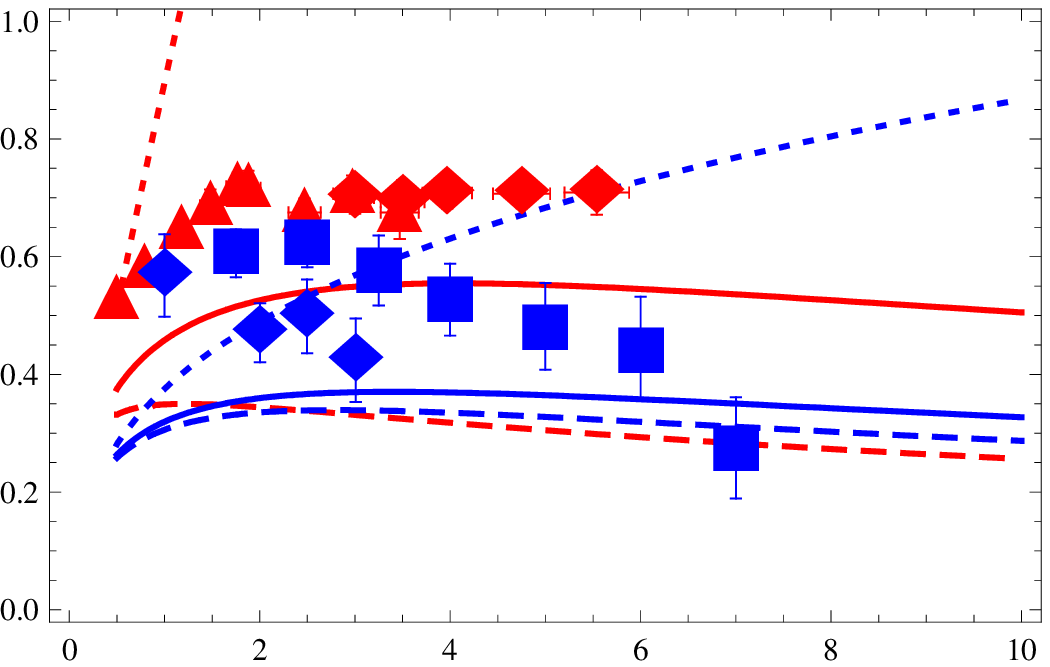}
\caption{LCSR results (solid curves) for the ratio  $\sqrt{Q^2} F_2^p/( F_1^p * 1.79)$ 
obtained using the BLW model (solid line), 
the asymptotic model (dashed line)
and the QCD sum rule model (dotted line) of the nucleon distribution amplitudes.
The corresponding results for the lattice values of the nucleon distribution amplitudes are given
in blue. 
{Red symbols}: experimental values obtained via Polarization transfer.
{Blue symbols}: experimental values obtained via Rosenbluth
separation. Currently the Rosenbluth separation data for $G_E$ are judged to be
less reliable.
} 
\label{fig:F2/F1}
\end{figure}
\\
To our accuracy, the sum rules for the nucleon form factors depend only on
the five parameters $f_N/\lambda_1$, $A_1^u$, $V_1^d$, $f_1^u$ and $f_1^d$.
Within the light-cone sum rule approach we determine the form factors $F_1$ and $F_2$
directly. The electric and the magnetic form factors  $G_E$ and $G_M$ are linear
combinations of $F_1$ and $F_2$:
\begin{align}
G_M (Q^2) & = F_1 (Q^2) + F_2 (Q^2) \, ,
\nonumber
\\
G_E (Q^2) & = F_1 (Q^2) - \frac{Q^2}{4 m_N^2} F_2 (Q^2) \, .
\label{defGEGM}
\end{align}
As discussed above,  Eq.~(\ref{defGEGM}) shows that in $G_E$ cancellations  occur. Therefore
our predictions for $G_E$ are less reliable than those for $G_M$.
\\
The light-cone sum rule predictions for the form factors are shown in Figs.~\ref{fig:GMGEp}~-~\ref{fig:F2/F1}.
For $G_M^p, G_A^p, G_M^n $ and $G_E^n$ the differences between the lattice determinations 
and the other approaches (asymptotic, QCD sum rule and BLW) are smaller than the expected overall uncertainties, 
i.e., the pairs of parameter sets (1) - (4), (2) - (5) and (3) - (6) yield almost identical results. 
Since $f_1^u$ and $f_1^d$ were chosen to be identical within these pairs, 
differences can only occur due to $V_1^d$, $A_1^u$ and $f_N/\lambda_1$. The lattice values
for $V_1^d$ and $A_1^u$ are very close to the BLW values, while  $f_N/\lambda_1$ is in the lattice determination 
about a factor of two smaller than the QCD sum rule estimate.
In $G_E$, $\sqrt{Q^2} F_2^p/( (\mu_p -1) F_1^p)$ and $G_T^p$ cancellations occur, so we expect
much bigger theoretical uncertainties and also big differences between our data sets might be possible.
\\
The data for $G_M^p$ (Fig.~\ref{fig:GMGEp}) are very well described with the asymptotic and the BLW data sets, the differences
between the parameter sets (2) and (5) and between (3) and (6) are negligible. The pure QCD sum rule estimates 
(set (1) and (4)) are about a factor of two too small. In the case of $G_M^n$ (Fig.~\ref{fig:GMGEn}) one sees the same structure 
for the different models of the nucleon distribution amplitude as for $G_M^p$, but now all theory 
predictions are shifted to lower values.
For $G_A^p$ (Fig.~\ref{fig:GAGT}) we agree for $Q^2$ values below 5 GeV$^2$ very well with the dipole behavior, if we use the
asymptotic or the BLW parameters; for higher $Q^2$ we predict a slightly steeper fall-off. 
Again the pure QCD sum rule estimates are considerably worse.
\\
An interesting test of our approach is whether the unphysical tensor form factor $G_T$  (Fig.~\ref{fig:GAGT}) 
is consistent 
with zero. This holds for all parameters sets except the pure QCD sum rule estimates (set (1)).
In our approach $G_T^p$ is not exactly zero, because we treat the intial proton state differently
from the final proton state: one is described by an interpolating nucleon field, the other by the
nucleon distribution amplitude.
\\
Finally we have the ratios $G_E^n/G_D$ (Fig.~\ref{fig:GMGEn}) , $G_E^p / G_M^p$ (Fig.~\ref{fig:GMGEp}),  
and  $\sqrt{Q^2} F_2^p/( (\mu_p-1) F_1^p )$ (Fig.~\ref{fig:F2/F1}), which
are very sensitive to the explicit form of the nucleon distribution amplitudes due
to cancellations in $G_E$.
If we just look at $G_E^n$ our result would be consistent with zero and therefore describes the data well.
If we investigate $G_E^n/G_D$, we blow up the large $Q^2$ contributions. Now we have a ``perfect'' agreement
between the pure QCD sum rule parameters and the data. Our data set (4) is almost identical  to the updated 
Galster fit.
The BLW model is consistent with zero, while the asymptotic distribution amplitude yields negative values.
The difference between the lattice values for the distribution amplitudes and the data sets (1) - (3) is 
visible, but not dramatic.
In the case of $G_E^p / G_M^p$,  and  $\sqrt{Q^2} F_2^p/( F_1^p * 1.79)$ we can make similar observations.
The purely asymptotic values lie above (below) the data for   $G_E^p / G_M^p$  ($\sqrt{Q^2} F_2^p/( F_1^p * 1.79)$),
the BLW data set moves the results in the right direction, but not far enough. Our data set (1) is completely off, because 
it predicts a very small value for $F_1^p$. Now we also have big differences between the lattice values of the 
distribution amplitudes and the pure QCD sum rule values.
\\
Taking into account that the full ${\cal{O}} (\alpha_s)$-corrections to the LCSRs are not yet available, already 
a rough agreement of our approach with the data is a success.
For $G_M^p, G_A^p, G_M^n $ and $G_E^n$ we get unexpectedly ``good'' results if we use the BLW form or the asymptotic
form of the nucleon distribution amplitude. The corresponding lattice values (set (5) and (6)) give similar results.
For $G_E^n/G_D$, $G_E^p / G_M^p$,  and  $\sqrt{Q^2} F_2^p/( F_1^p * 1.79)$ the BLW model lies in the right 
ball park, but we would prefer a model for the distribution amplitudes which lies in the middle between
the asymptotic and the pure QCD sum rule estimate (BLW is closer to the asymptotic value).
\\
In the transition form factors of $\gamma^* N \to \Delta$ all eight non-perturbative parameters 
appear, see \cite{BLPR2005}.
The results are shown in Fig.~\ref{fig:Delta}. As expected, now the differences between
the parameter set pairs (1) - (4), (2) - (5) and (3) - (6) are more pronounced.
In the case of $G_M^*$ the theory curves generally tend to be more flat than the 
experimental data. 
The form factors obtained with the lattice values for the nucleon distribution amplitude 
lie considerably above the data sets (1) - (3). 
Above $Q^2 \approx$ 3 GeV$^2$ the asymptotic distribution amplitude and the 
BLW distribution amplitude are close to the data. The fact that
$R_{EM}$ is close to zero is reproduced very well with the BLW parameters
(sets (3) and (6)) and the lattice plus asymptotic values (set(5)), 
while positive values are obtained with the purely asymptotic form (set (2)). One gets a negative result 
with the QCD sum rule determination of the nucleon distribution amplitude  (sets (1) and (4)).
In the case of $R_{SM}$ the differences are not very pronounced, all values are close to zero.
Altogether one has to conclude that while all approaches give the correct order of magnitude 
none gives a really convincing description of all data. However in view of the fact that the systematic 
uncertainities are even more pronounced than for the nucleon form factors more could not have
been expected.
\section{Reducing the number of independent parameters of the distribution amplitudes}
In \cite{BMR2008} the following approximate relations between twist-4 and twist-3 parameters
were derived:
\begin{eqnarray}
f_1^d &=& \frac{3}{10} - \frac16 \frac{f_N}{\lambda_1}\, ,
\nonumber
\\                                        
f_1^u &=& \frac{1}{10} - \frac16 \frac{f_N}{\lambda_1}\, . 
\label{integrable}
\end{eqnarray}
Using these relations, we can express the nucleon form factors in terms of only three independent parameters,
namely $\frac{f_N}{\lambda_1}$, $V_1^d$ and $A_1^u$. 
For the comparison with the data we show now only two models for the remaining three parameters of the nucleon 
distribution amplitude:
\begin{itemize}
\item[(a)] Lattice determination of the distribution amplitude - blue curve.
\item[(b)] BLW model - red curve.
\end{itemize}
\begin{figure*}
       \centering
       \subfigure{\includegraphics[width=0.47\textwidth,clip]{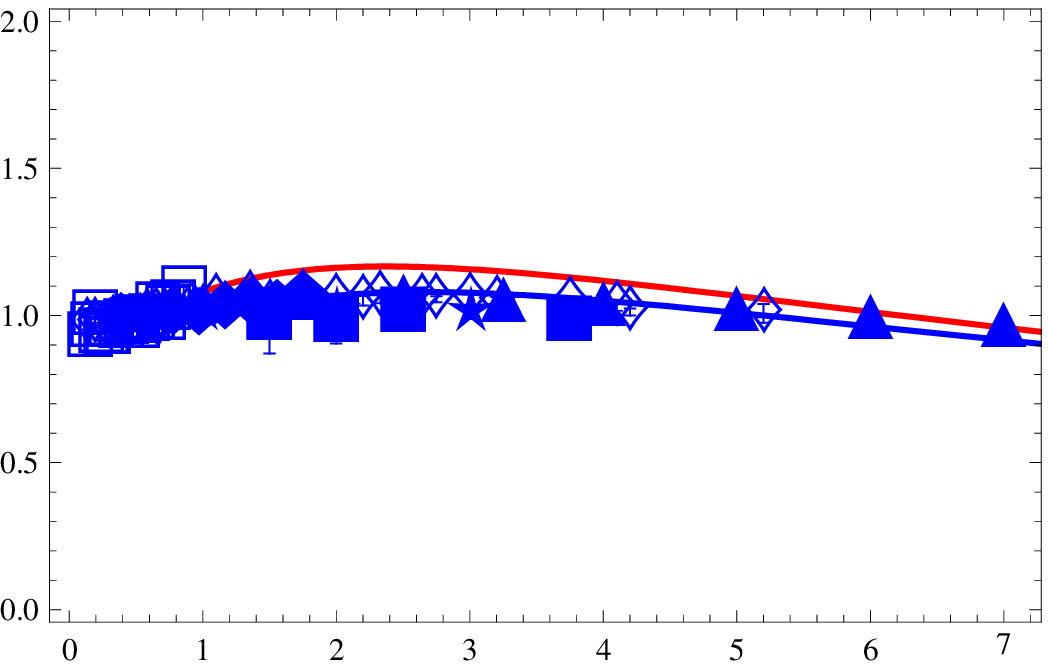}}
       \hspace{0.03\textwidth}
       \subfigure{\includegraphics[width=0.47\textwidth,clip]{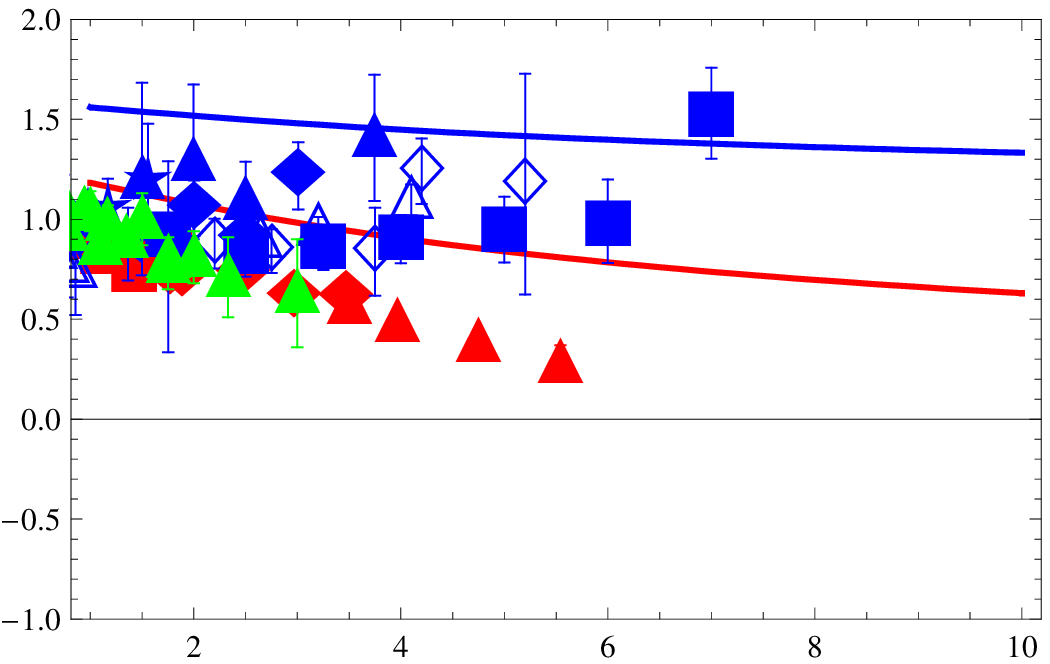}}
\caption{LCSR results for the electromagnetic form factors 
(left: $G_M/( \mu_p G_D)$ vs. $Q^2$; right:  $ \mu_p G_E/G_M$ vs. $Q^2$) of the 
proton, obtained using the BLW model (red solid line) and the lattice prediction 
(blue solid line) for the nucleon distribution amplitudes. In both cases the parameters $f_1^x$ 
are determined from the twist-3 parameters, cf. Eq. (47).
The red data points on the right picture are JLAB data, while the blue and the green ones
are obtained via Rosenbluth separation.
}
\label{fig:GMGEpI}
\end{figure*}
\begin{figure*}
\centering
\subfigure{\includegraphics[width=0.47\textwidth,clip]{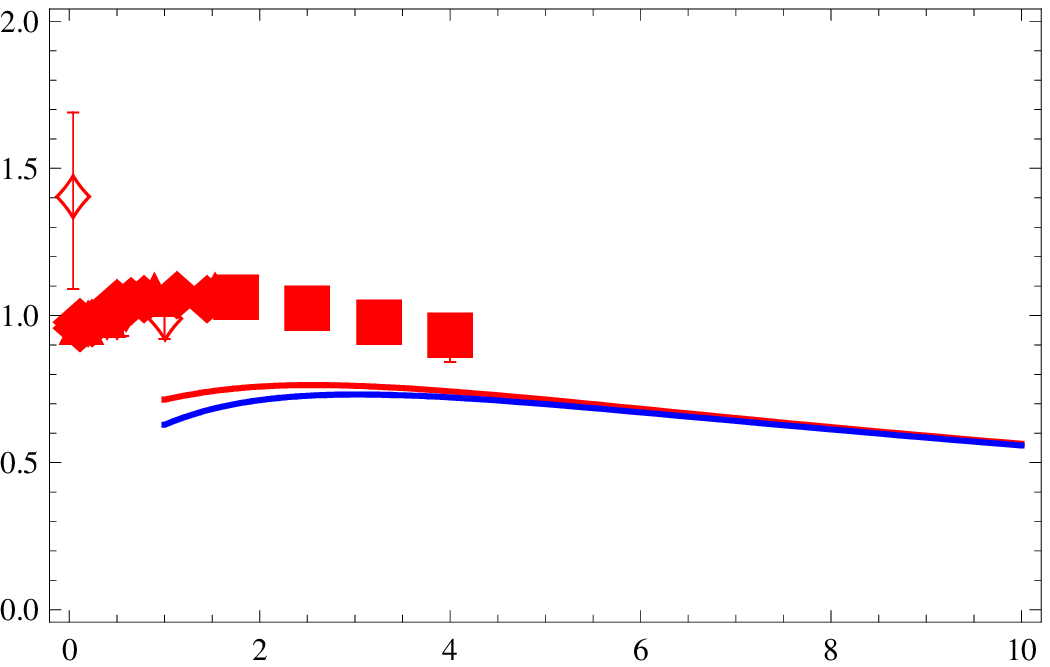}}
\hspace{0.03\textwidth}
\subfigure{\includegraphics[width=0.47\textwidth,clip]{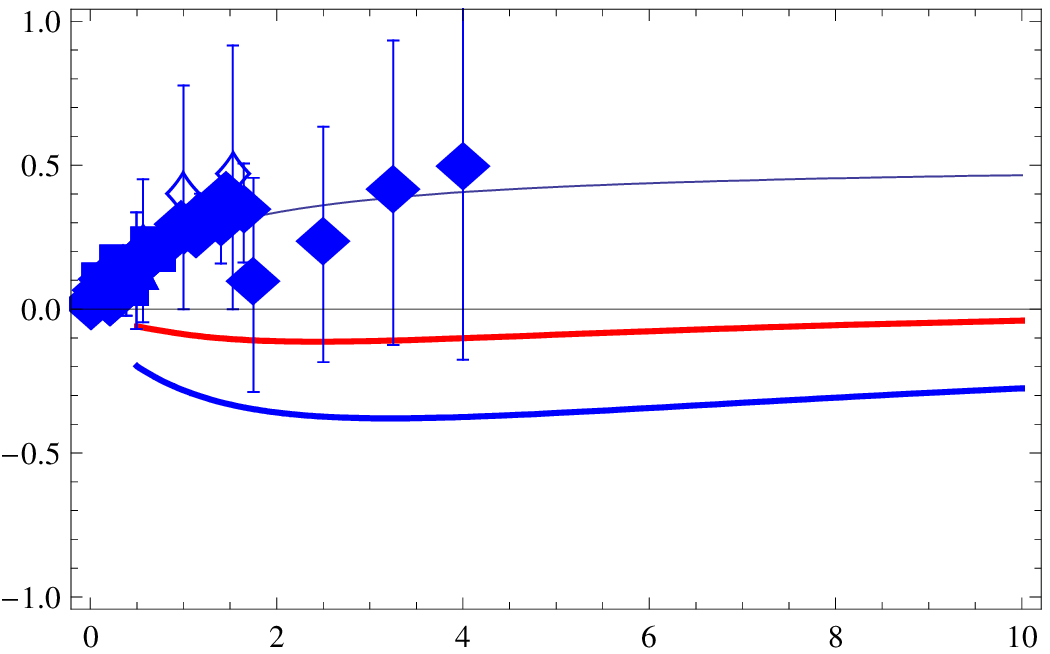}}
\caption{LCSR results for the electromagnetic form factors of the 
neutron (left: $G_M/(\mu_n G_D)$ vs. $Q^2$; right:  $G_E/(G_D)$ vs. $Q^2$), 
obtained using the BLW model (red solid line) 
and the lattice prediction 
(blue solid line) for the nucleon distribution amplitudes. In both cases the parameters $f_1^x$ 
are determined from the twist-3 parameters, cf. Eq. (47).
The thin solid blue line represents the updated Galster fit.
}
\label{fig:GMGEnI}
\end{figure*}
\begin{figure*}
       \centering
       \subfigure{\includegraphics[width=0.47\textwidth,clip]{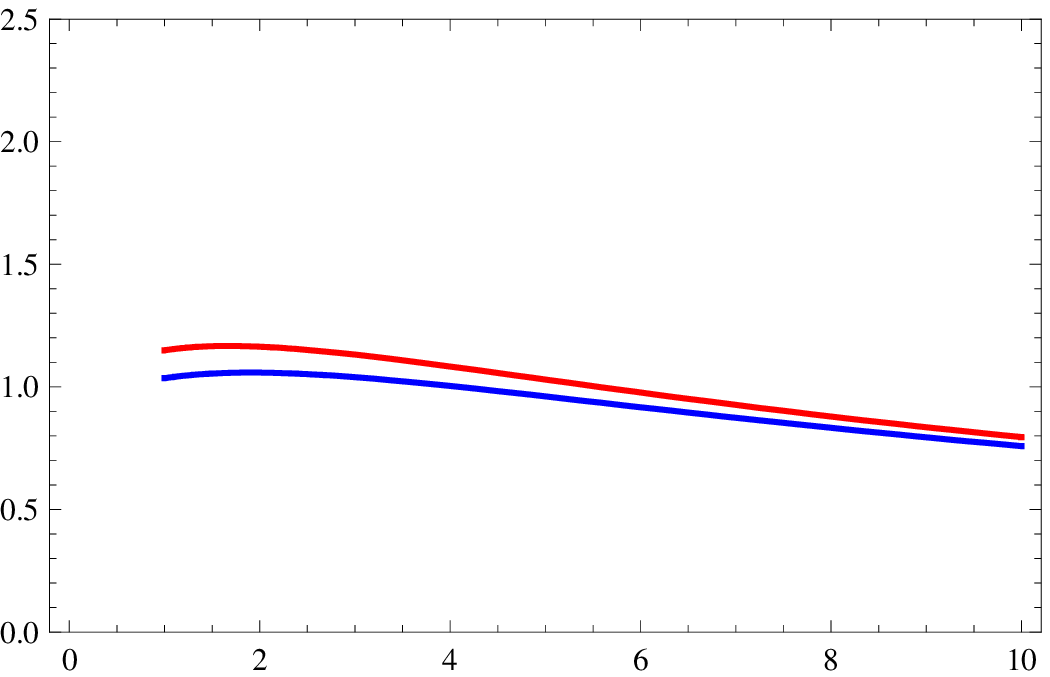}}
       \hspace{0.03\textwidth}
       \subfigure{\includegraphics[width=0.47\textwidth,clip]{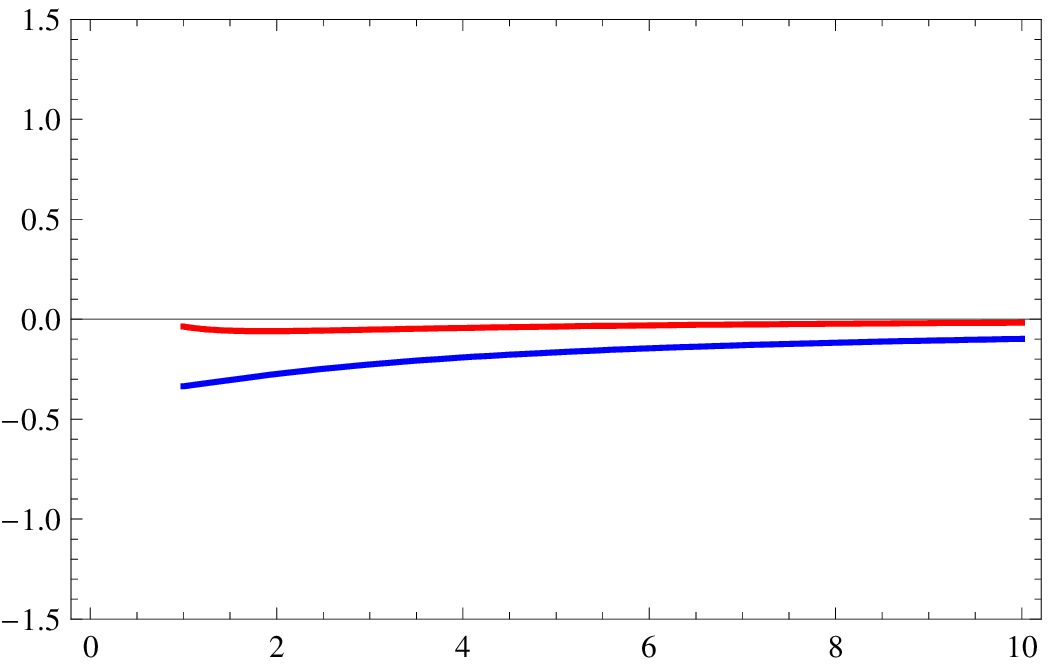}}
\caption{LCSR results (solid curves) for the axial form factor  of the proton
$G_A$ normalized to $G_D=g_A/(1+Q^2/m_A^2)^2$  vs. $Q^2$(left panel) and 
the tensor form factor $G_T$ normalized to $G_A$  vs. $Q^2$ (right panel),
obtained using the BLW model (red solid line) 
and the lattice prediction 
(blue solid line) for the nucleon distribution amplitudes. In both cases the parameters $f_1^x$ 
are determined from the twist-3 parameters, cf. Eq. (47).
}
\label{fig:GAGTI}
\end{figure*} 
\begin{figure}[ht]
\centering
  \includegraphics[width=0.8\textwidth,angle=0]{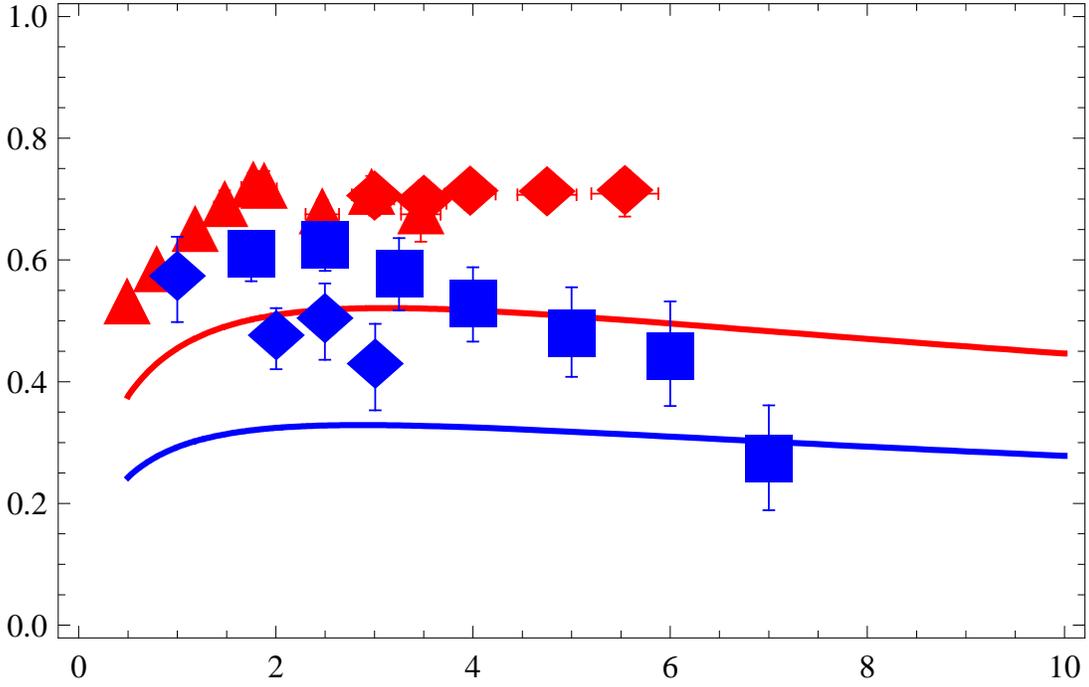}
\caption{LCSR results (solid curves) for the ratio  $\sqrt{Q^2} F_2^p/( F_1^p * 1.79)$ 
obtained using the BLW model (red solid line) 
and the lattice prediction 
(blue solid line) for the nucleon distribution amplitudes. In both cases the parameters $f_1^x$ 
are determined from the twist-3 parameters, cf. Eq. (47).
{ Red symbols}: experimental values obtained via Polarization transfer.  
{Blue symbols}: experimental values obtained via Rosenbluth
separation.
} 
\label{fig:F2/F1I}
\end{figure}
\begin{figure*}[ht]
\centering
       \subfigure{\includegraphics[width=0.315\textwidth,clip]{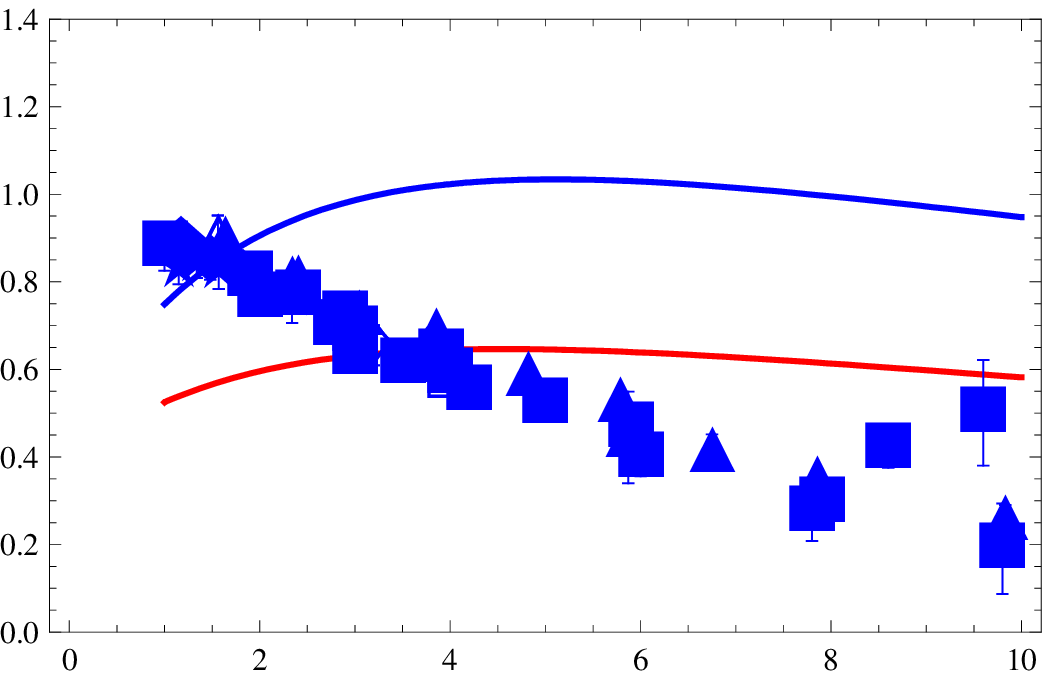}}
       \hspace{0.01\textwidth}
       \subfigure{\includegraphics[width=0.315\textwidth,clip]{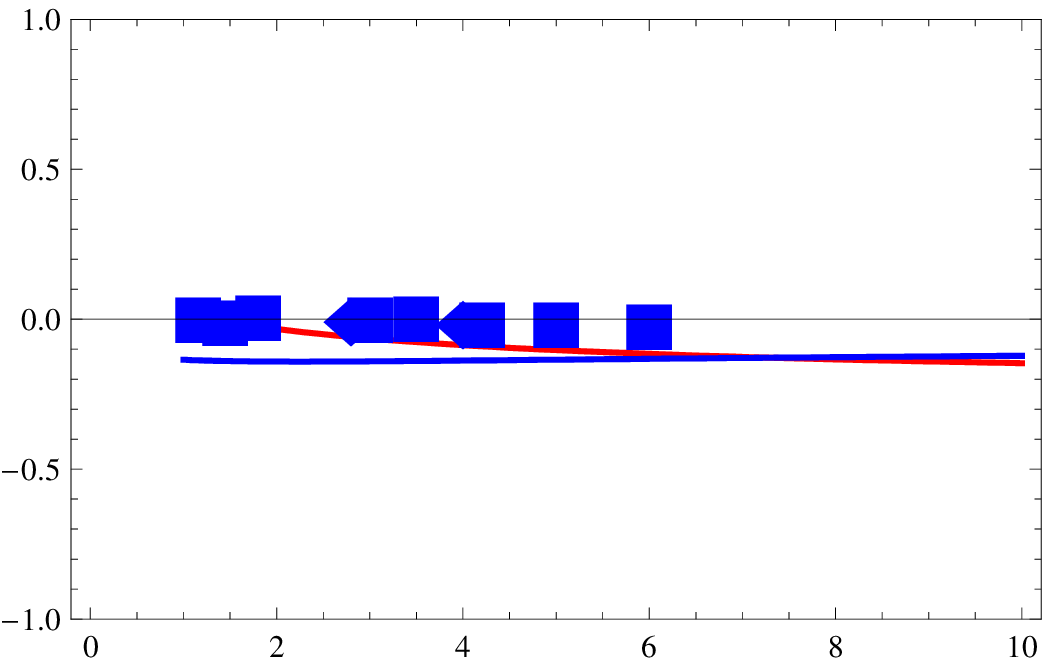}}
       \hspace{0.01\textwidth}
       \subfigure{\includegraphics[width=0.315\textwidth,clip]{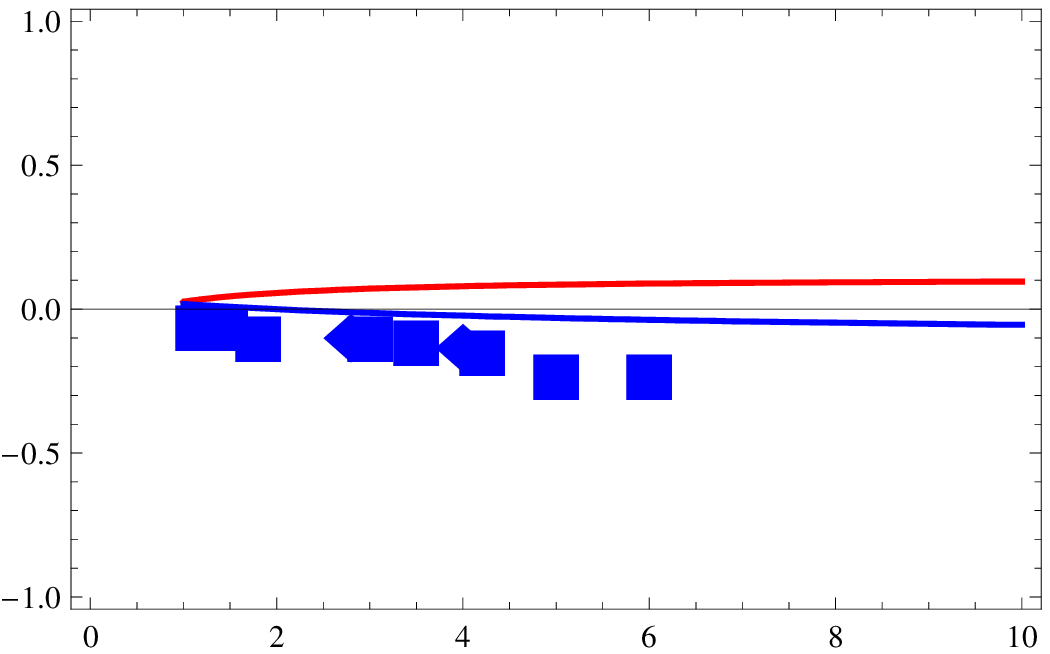}}
\caption{$\gamma^* N \to \Delta$ transition form factors 
(left:$G_M^*/(3 G_D)$ vs. $Q^2$, middle:$R_{EM}$  vs. $Q^2$, right: $R_{SM}$  vs. $Q^2$)
in  the LCSR approach~\cite{BLPR2005}
obtained using the BLW model (red solid line)
and the lattice prediction 
(blue solid line) for the nucleon distribution amplitudes. In both cases the parameters $f_1^x$ 
are determined from the twist-3 parameters, cf. Eq. (47).
}
\label{fig:DeltaI}
\end{figure*}
We obtain the following values for $f_1^x$:
\begin{displaymath}
\begin{array}{|c||c|c||c|c|c|}
\hline
      & (a)  & (b) & \mbox{asymptotic} & \mbox{BLW} & \mbox{QCD-SR}
\\
\hline
\hline
f_1^u &  0.11 &  0.13   &   0.10             & 0.09        &  0.07
\\
\hline
f_1^d &  0.31 &  0.33  &   0.30             &  0.33       &  0.40
\\
\hline
\end{array}
\end{displaymath}
which are compared with our previous estimates for $f_1^u$ and $f_1^d$. In this approach
$f_1^d$ lies between the asymptotic and the BLW value; $f_1^u$ is also close to the asymptotic
or the BLW value, but its deviation from the asymptotic value is in the ``wrong'' direction.
In Fig.~\ref{fig:GMGEpI} we show the electromagnetic form factors of the proton,
in Fig.~\ref{fig:GMGEnI} the electromagnetic form factors of the neutron,
in  Fig.~\ref{fig:GAGTI} the axial and the tensor form factor of the proton and finally in
Fig.~\ref{fig:F2/F1I} the ratio of the form factors $F_2$ and $F_1$ of the proton.
In Fig.~\ref{fig:DeltaI} we show the three $N \to \Delta$ transition form factors.
In all cases we obtain results which are very close the BLW results of the previous section, 
so it seems that the nucleon form factors are very sensitive to the values of $V_1^d$ and $A_1^u$, 
while the dependence on $f_N/\lambda_1$, $f_1^d$ and $f_1^u$ is less pronounced.
\section{Effects of higher conformal spin contributions}
In this section we include (in comparison to the previous sections) also contributions
of the next-to-next-to leading conformal spin to the leading-twist distribution amplitude.
These terms have been determined on the lattice \cite{Lattice2007,Lattice2008} and with QCD sum rules 
\cite{Chernyak1984,King1987,COZ1987}. 
The explicit expressions for the leading-twist distribution amplitudes can be found in appendix C.
The contributions of these higher moments to the electromagnetic form factors 
of the nucleon have already been estimated in \cite{BLMS2001}, but only for the Chernyak-Zhitnitsky
interpolating field. Here we work out the contributions of the second moments to the
light-cone sum rules for nucleon form factors using the Ioffe current.
We will use the form in Eq.~(\ref{tw3dwave}) for the leading-twist distribution amplitude and 
the following parameter sets:
\begin{enumerate}
\item Asymptotic distribution amplitude                         (black       lines).
\item BLW plus second moments from QCD sum rules                (dotted red  lines).
\item BLW plus second moments from the lattice                  (dashed red  lines).
\item BLW plus second moments \'a la BLW                        (solid  red  lines).
\item Lattice evaluation plus QCD sum rule estimates for $f_x^y$ (dotted blue lines).
\item Lattice evaluation plus asymptotic values     for $f_x^y$ (dashed blue lines).
\item Lattice evaluation plus BLW          estimates for $f_x^y$ (solid  blue lines).
\end{enumerate}
\begin{figure*}
\centering
       \subfigure{\includegraphics[width=0.47\textwidth,clip]{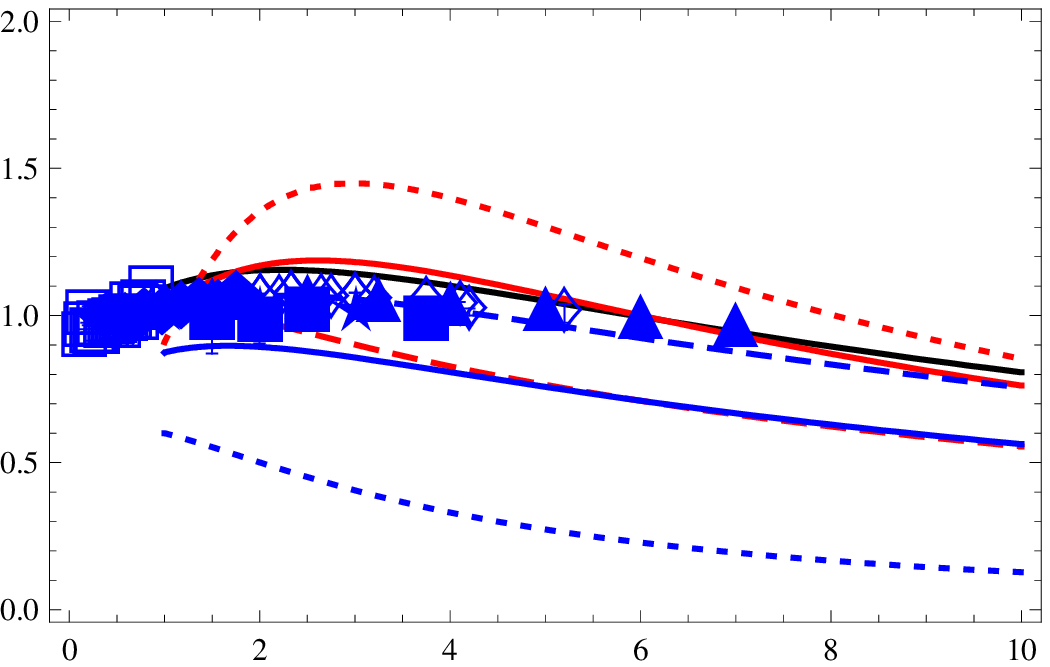}}
       \hspace{0.03\textwidth}
       \subfigure{\includegraphics[width=0.47\textwidth,clip]{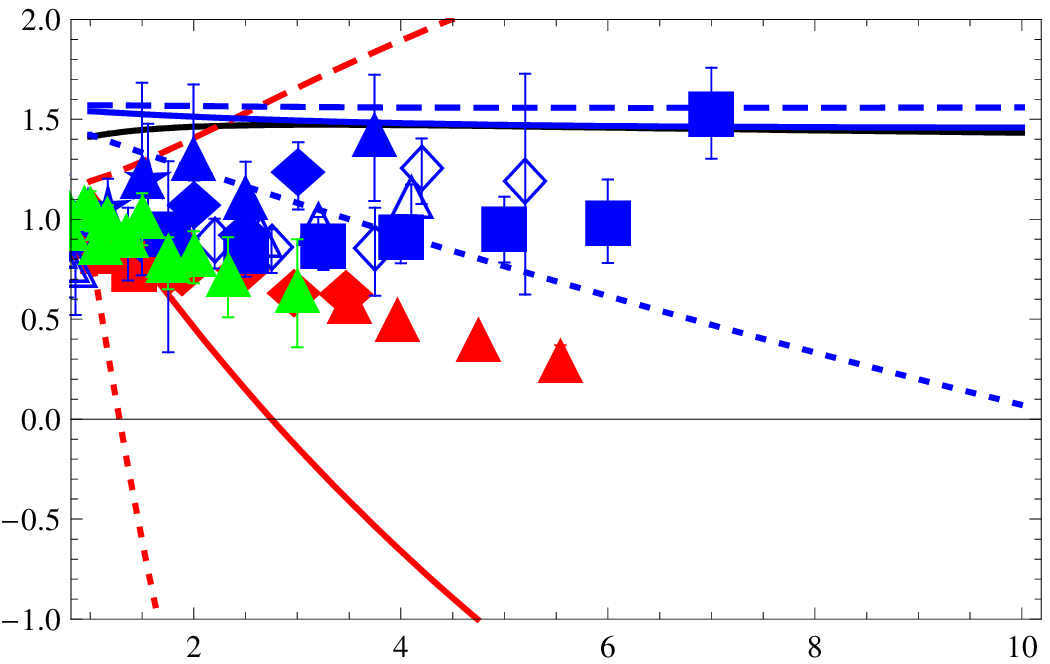}}
\caption{LCSR results for the electromagnetic form factors 
(left: $G_M/( \mu_p G_D)$ vs. $Q^2$; right:  $ \mu_p G_E/G_M$ vs. $Q^2$) of the 
proton, 
obtained using the asymptotic form (black solid line), the BLW model (red) 
with second moments from QCD sum rules (dotted), lattice (dashed) and BLW (solid)
and the lattice determination (blue) with $f_x^y$ from QCD sum rules (dotted), 
the asymptotic model (dashed) and the BLW model (solid).
The red data points on the right picture are JLAB data, while the blue and green ones
are obtained via Rosenbluth separation.
}
\label{fig:GMGEpd}
\end{figure*}
\begin{figure*}
\centering
       \subfigure{\includegraphics[width=0.47\textwidth,angle=0]{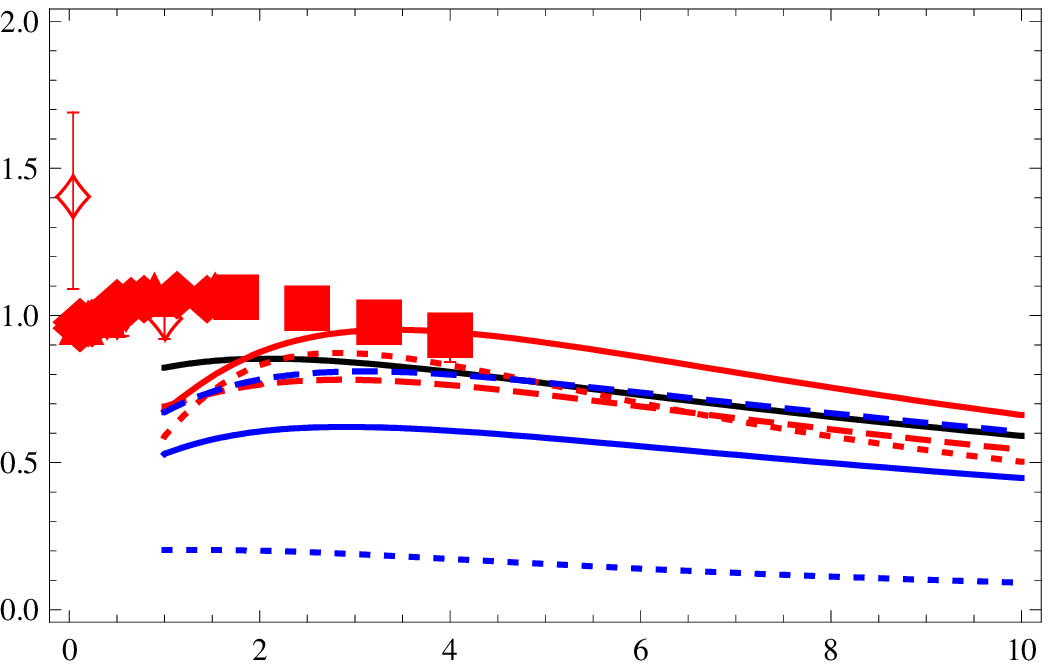}}
       \hspace{0.03\textwidth}
       \subfigure{\includegraphics[width=0.47\textwidth,angle=0]{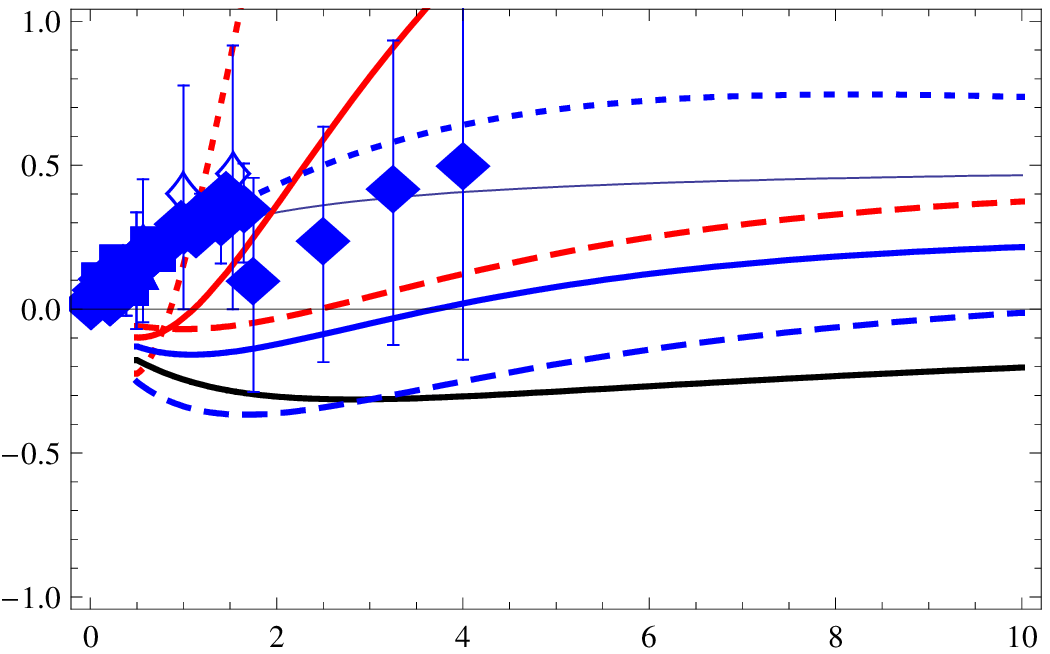}}
\caption{LCSR results for the electromagnetic form factors of the 
neutron (left: $G_M/(\mu_n G_D)$ vs. $Q^2$; right:  $G_E$ vs. $Q^2$), 
obtained using the asymptotic form (black solid line), the BLW model (red) 
with second moments from QCD sum rules (dotted), lattice (dashed) and BLW (solid)
and the lattice determination (blue) with $f_x^y$ from QCD sum rules (dotted), 
the asymptotic model (dashed) and the BLW model (solid).
}
\label{fig:GMGEnd}
\end{figure*}
\begin{figure*}
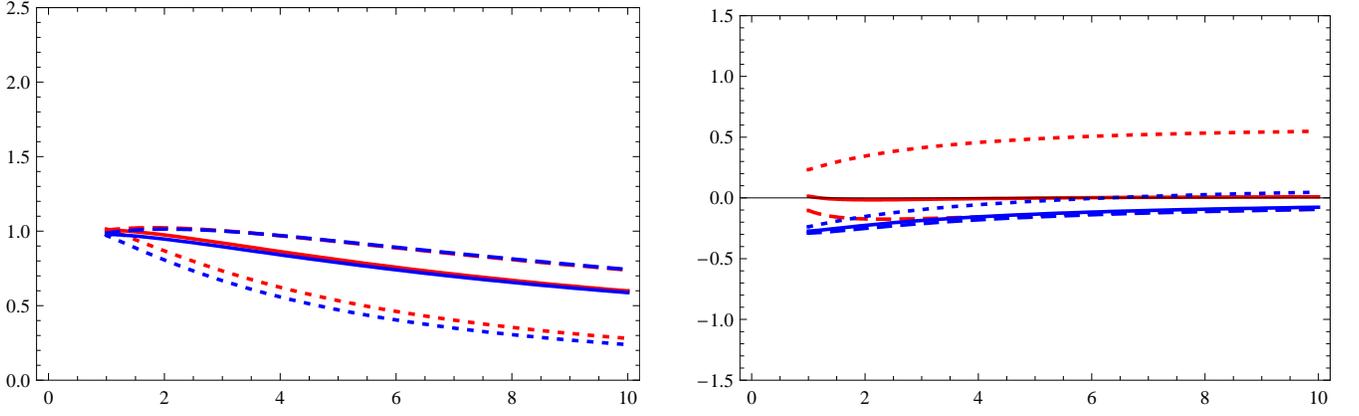

\centering
       \subfigure{\includegraphics[width=0.47\textwidth,angle=0]{GA08}}
       \hspace{0.03\textwidth}
       \subfigure{\includegraphics[width=0.47\textwidth,angle=0]{GToverGA08}}
\caption{LCSR results (solid curves) for the axial form factor  of the proton
$G_A$ normalized to $G_D=g_A/(1+Q^2/m_A^2)^2$  vs. $Q^2$(left panel) and 
the tensor form factor $G_T$ normalized to $G_A$  vs. $Q^2$ (right panel),
obtained using the asymptotic form (black solid line), the BLW model (red) 
with second moments from QCD sum rules (dotted), lattice (dashed) and BLW (solid)
and the lattice determination (blue) with $f_x^y$ from QCD sum rules (dotted), 
the asymptotic model (dashed) and the BLW model (solid).
}
\label{fig:GAGTd}
\end{figure*} 
\begin{figure*}
\centering
  \includegraphics[width=0.45\textwidth,clip]{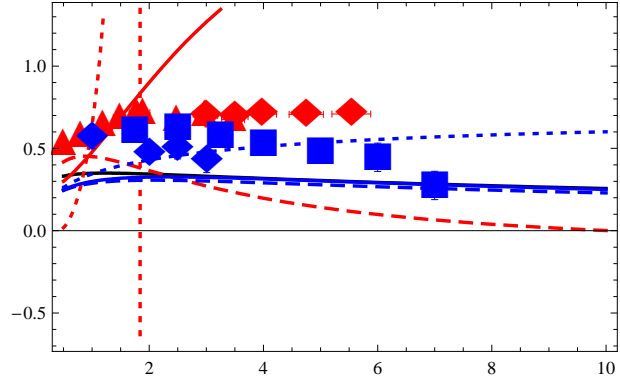}
\caption{LCSR results (solid curves) for the ratio  $\sqrt{Q^2} F_2^p/( F_1^p * 1.79)$ 
obtained using the asymptotic form (black solid line), the BLW model (red) 
with second moments from QCD sum rules (dotted), lattice (dashed) and BLW (solid)
and the lattice determination (blue) with $f_x^y$ from QCD sum rules (dotted), 
the asymptotic model (dashed) and the BLW model (solid).
{ Red symbols}: experimental values obtained via Polarization transfer.  
{Blue symbols}: experimental values obtained via Rosenbluth
separation.
} 
\label{fig:F2/F1d}
\end{figure*}
\begin{figure*}
\centering
       \subfigure{\includegraphics[width=0.315\textwidth,clip]{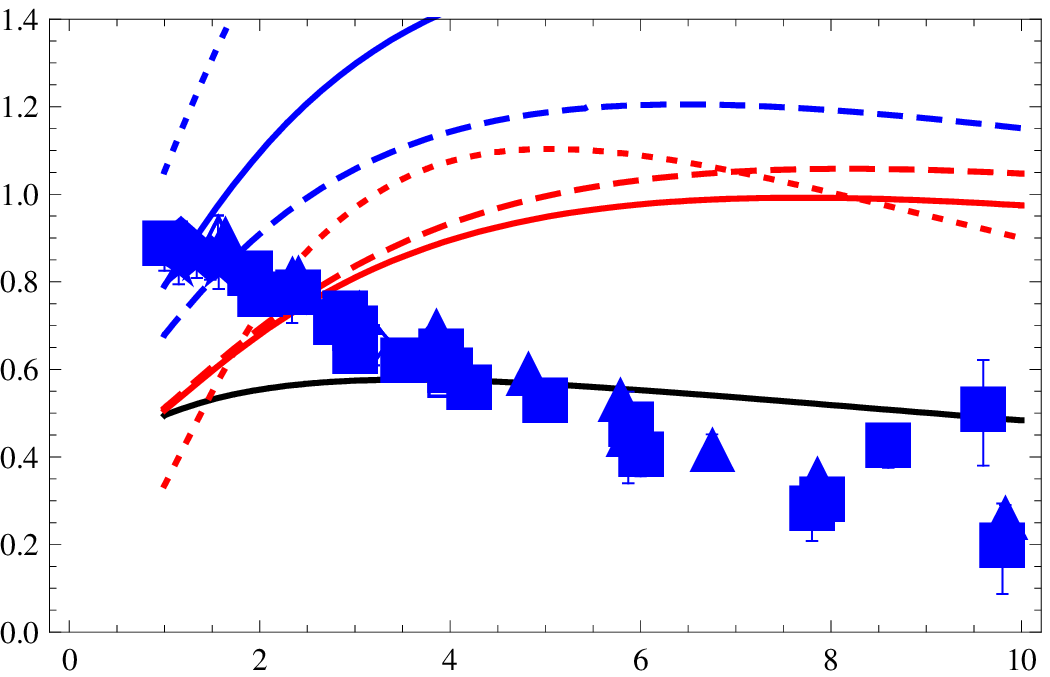}}
       \hspace{0.01\textwidth}
       \subfigure{\includegraphics[width=0.315\textwidth,clip]{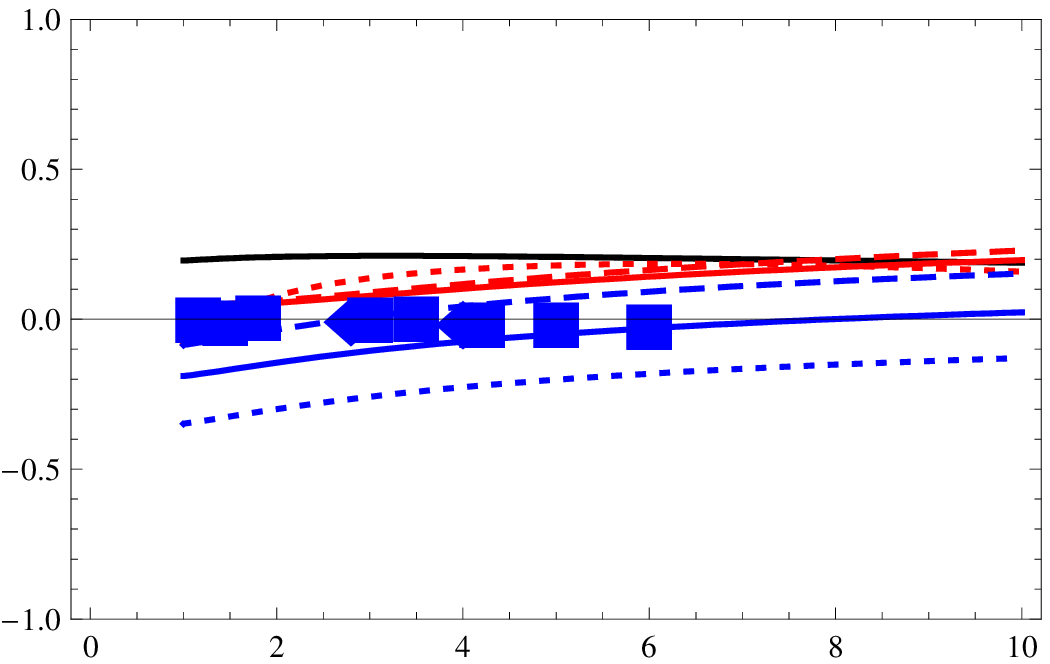}}
       \hspace{0.01\textwidth}
       \subfigure{\includegraphics[width=0.315\textwidth,clip]{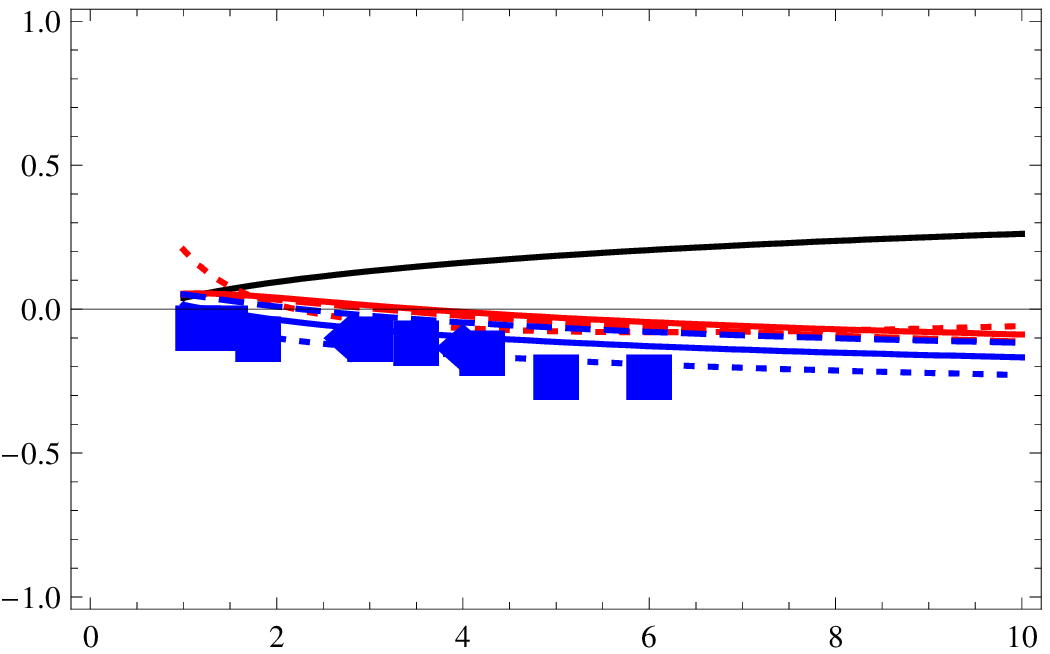}}
\caption{$\gamma^* N \to \Delta$ transition form factors 
(left:$G_M^*/(3 G_D)$ vs. $Q^2$, middle:$R_{EM}$  vs. $Q^2$, right: $R_{SM}$  vs. $Q^2$)
in  the LCSR approach~\cite{BLPR2005}
obtained using the asymptotic form (black solid line), the BLW model (red) 
with second moments from QCD sum rules (dotted), lattice (dashed) and BLW (solid)
and the lattice determination (blue) with $f_x^y$ from QCD sum rules (dotted), 
the asymptotic model (dashed) and the BLW model (solid).
}
\label{fig:Deltad}
\end{figure*}
In Fig.~\ref{fig:GMGEpd} we show the electromagnetic form factors of the proton,
in Fig.~\ref{fig:GMGEnd} the electromagnetic form factors of the neutron,
in  Fig.~\ref{fig:GAGTd} the axial and the tensor form factor of the proton and finally in
Fig.~\ref{fig:F2/F1d} the ratio of the form factors $F_2$ and $F_1$ of the proton.
In Fig.~\ref{fig:Deltad} we show the three $N \to \Delta$ transition form factors.
\\
In almost all cases the second moments of the leading-twist distribution 
amplitude determined with QCD sum rules give huge corrections. We show these parameter sets in the plots, 
but we will not discuss them any further.
\\
The magnetic form factor of the proton $G_M^p$ is very well described by the 
BLW model with second moments \'a la BLW (set(4)) or from the lattice (set(3)) 
and the lattice values for the distribution amplitude with
$f_x^y$ from BLW (set (7)) or with the asymptotic values for $f_x^y$ (set (6)).
This is not unexpected, since the second moments \'a la BLW and from the lattice are quite
similar in size. This observation ensures, however, that there is not an unexpected strong 
sensitivity of the LCSRs to the second moments.
The theory predictions for the magnetic form factor of the neutron $G_M^n$ are again shifted
to lower values. Apart from this fact, the predictions for the parameter sets 
(1), (3), (4), (6), (7) lie relatively close together and they agree a little
better with experiment, compared to the case where the $d$-wave contributions
have been neglected.
Also for $G_A^p$ and $G_T^p$ we get nice results, unless we use the QCD sum rule values of the second moments.
As expected, in $G_E^p/G_M^p$, $G_E^n/G_D$ and $F_2^p/F_1^p$ cancellations arise that lead to a strong
dependence on the concrete form of the nucleon distribution amplitudes.
\\
In the case of the $N \to \Delta$ transition the inclusion of $d$-wave corrections leads to strong 
enhancements in the prediction of $G_M^*$, while $R_{EM}$ and $R_{SM}$ agree now better with experiment.
\section{Conclusion}
We have compared a new determination of the nucleon distribution amplitudes
based on  lattice QCD with different values available in the literature.
The non-perturbative parameters of non-leading conformal spin from the lattice evaluation turned out to be
close to the asymptotic form and very close to the BLW model.
For the leading conformal spin parameters $f_N$, $\lambda_1$ and $\lambda_2$ the deviation between
lattice and QCD sum rules is about $30\%$, which is possibly due to neglected radiative corrections 
in the QCD sum rules estimates.
Our models for the nucleon distribution amplitudes can be related to measurable
form factors with light-cone sum rules. Despite the fact that the light-cone sum rules
are only calculated to leading order in QCD and despite an intrinsic uncertainty
of light-cone sum rules of about $\pm 20 \%$ we get a very good 
description of $G_M^p$, $G_M^n$, $G_A^p$ and $G_T^p$ at intermediate momentum transfer.
In  $G_E^p$, $G_E^n$ and $F_2^p/F_1^p$ cancellations occur, which limit our 
predictive power.
In general we found the following tendency:
The asymptotic distribution amplitudes describe the data already amazingly well.
Pure QCD sum rule estimates for the non-perturbative parameters overestimate the 
deviation from the asymptotic form, but the deviation goes in the right direction.
The best results are obtained for the BLW values and the very similar lattice values.
Including also $d$-wave contributions to the twist-3 distribution amplitude improves  
the description of $G_M^p$, $G_M^n$, $G_A^p$ and $G_T^p$  a little bit,
but results also in bigger uncertainites in $G_E^p$, $G_E^n$ and $F_2^p/F_1^p$.
In the case of the $N \to \Delta$ transition, we do not see the steep fall-off of $G_M^*$, 
but the smallness of $R_{EM}$ and $R_{SM}$ is very well reproduced.
\\
Further improvements on the theoretical side can be achieved by determining the
NLO QCD corrections to light-cone sum rules, which connect the nucleon distribution
amplitudes to the form factors. To match the NLO QCD accuracy also $\alpha_s$-corrections
have then to be included in all QCD sum rule estimates of the moments of the nucleon
distribution amplitudes.
\section{Acknowledgment}
We would like to thank V. Braun and A. Sch{\"a}fer for discussions and comments on the manuscript, 
R. Schiavilla for providing the data for $G_E^n$ from Ref. \cite{ExpGEn16} and
Y. Aoki for clarifying comments. All lattice data were provided by the QCDSF collaboration. This work was supported
by DFG (Forschergruppe Gitter-Hadronen-Ph{\"a}nomenologie and SFB/TR55 Hadron Physics from Lattice QCD)
and by BMBF.

\appendix
\section{Nucleon distribution amplitudes up to twist-6}
For completeness we give in this appendix the full expressions for
the nucleon distribution amplitudes up to twist-6, details can be
found in \cite{BFMS2000,BLW2006}.
The general Lorentz decomposition of the matrix element defined in Eq.~(\ref{def1-NDA})
reads \cite{BFMS2000}
\begin{eqnarray}
\label{fullDA}
\lefteqn{ 4 \bra{0} \epsilon^{ijk} u_\alpha^i(a_1 x) u_\beta^j(a_2 x) d_\gamma^k(a_3 x) \ket{P} }
\nonumber \\
&=&
{\cal S}_1 m_N C_{\alpha \beta} \left(\gamma_5 N\right)_\gamma +
{\cal S}_2 m_N^2 C_{\alpha \beta} \left(\!\not\!{x} \gamma_5 N\right)_\gamma +
{\cal P}_1 m_N \left(\gamma_5 C\right)_{\alpha \beta} N_\gamma +
{\cal P}_2 m_N^2 \left(\gamma_5 C \right)_{\alpha \beta} \left(\!\not\!{x} N\right)_\gamma
\nonumber \\
&& +
\left(\mathcal{V}_1+\frac{x^2m_N^2}{4}\mathcal{V}_1^M \right)
\left(\!\not\!{P}C \right)_{\alpha \beta} \left(\gamma_5 N\right)_\gamma +
{\cal V}_2 m_N \left(\!\not\!{P} C \right)_{\alpha \beta} \left(\!\not\!{x} \gamma_5 N\right)_\gamma  +
{\cal V}_3 m_N  \left(\gamma_\mu C \right)_{\alpha \beta}\left(\gamma^{\mu} \gamma_5 N\right)_\gamma
\nonumber \\
&& +
{\cal V}_4 m_N^2 \left(\!\not\!{x}C \right)_{\alpha \beta} \left(\gamma_5 N\right)_\gamma +
{\cal V}_5 m_N^2 \left(\gamma_\mu C \right)_{\alpha \beta} \left(i \sigma^{\mu\nu} x_\nu \gamma_5
N\right)_\gamma
+ {\cal V}_6 m_N^3 \left(\!\not\!{x} C \right)_{\alpha \beta} \left(\!\not\!{x} \gamma_5 N\right)_\gamma
\nonumber \\
&& +
\left(\mathcal{A}_1+\frac{x^2m_N^2}{4}\mathcal{A}_1^M\right)
\left(\!\not\!{P}\gamma_5 C \right)_{\alpha \beta} N_\gamma +
{\cal A}_2 m_N \left(\!\not\!{P}\gamma_5 C \right)_{\alpha \beta} \left(\!\not\!{x} N\right)_\gamma  +
{\cal A}_3 m_N  \left(\gamma_\mu \gamma_5 C \right)_{\alpha \beta}\left( \gamma^{\mu} N\right)_\gamma
\nonumber \\
&& +
{\cal A}_4 m_N^2 \left(\!\not\!{x} \gamma_5 C \right)_{\alpha \beta} N_\gamma +
{\cal A}_5 m_N^2 \left(\gamma_\mu \gamma_5 C \right)_{\alpha \beta} \left(i \sigma^{\mu\nu} x_\nu
N\right)_\gamma
+ {\cal A}_6 m_N^3 \left(\!\not\!{x} \gamma_5 C \right)_{\alpha \beta} \left(\!\not\!{x} N\right)_\gamma
\nonumber \\
&& +
\left(\mathcal{T}_1+\frac{x^2m_N^2}{4}\mathcal{T}_1^M\right)
\left(P^\nu i \sigma_{\mu\nu} C\right)_{\alpha \beta} \left(\gamma^\mu\gamma_5 N\right)_\gamma
+
{\cal T}_2 m_N \left(x^\mu P^\nu i \sigma_{\mu\nu} C\right)_{\alpha \beta} \left(\gamma_5 N\right)_\gamma
\nonumber \\
&&
+ {\cal T}_3 m_N \left(\sigma_{\mu\nu} C\right)_{\alpha \beta} \left(\sigma^{\mu\nu}\gamma_5 N\right)_\gamma
+ {\cal T}_4 m_N \left(P^\nu \sigma_{\mu\nu} C\right)_{\alpha \beta} \left(\sigma^{\mu\rho} x_\rho \gamma_5 N\right)_\gamma
\nonumber \\
&&
+ {\cal T}_5 m_N^2 \left(x^\nu i \sigma_{\mu\nu} C\right)_{\alpha \beta} \left(\gamma^\mu\gamma_5 N\right)_\gamma
+ {\cal T}_6 m_N^2 \left(x^\mu P^\nu i \sigma_{\mu\nu} C\right)_{\alpha \beta} \left(\!\not\!{x} \gamma_5 N\right)_\gamma
\nonumber 
\\&&
+ {\cal T}_{7} m_N^2 \left(\sigma_{\mu\nu} C\right)_{\alpha \beta} \left(\sigma^{\mu\nu} \!\not\!{x} \gamma_5 N\right)_\gamma
+ {\cal T}_{8} m_N^3 \left(x^\nu \sigma_{\mu\nu} C\right)_{\alpha \beta} \left(\sigma^{\mu\rho} x_\rho \gamma_5 N\right)_\gamma \,,
\nonumber \\
\end{eqnarray}
with

\begin{eqnarray}
{\cal S}_1 = S_1 \,,  && 2 (P \! \cdot \! x) \, {\cal S}_2 = S_1-S_2 \,, 
\nonumber 
\\
{\cal P}_1 = P_1 \,,  && 2 (P \! \cdot \! x) \, {\cal P}_2 = P_2-P_1 \,,
\nonumber 
\\
{\cal V}_1 = V_1 \,,  & \quad & 2 (P \! \cdot \! x) \, {\cal V}_2 = V_1 - V_2 - V_3  \,, 
\nonumber
\\
2 {\cal V}_3 = V_3 \,, && 4 (P \! \cdot \! x) \, {\cal V}_4  = - 2 V_1 + V_3 + V_4  + 2 V_5\,,
\nonumber
\\
4 (P \! \cdot \! x) {\cal V}_5 = V_4 - V_3\,, && 4 \left(P \! \cdot \! x\right)^2  {\cal V}_6  = - V_1 + V_2 +  V_3 +  V_4 + V_5 - V_6\,,\quad
\nonumber 
\\
{\cal A}_1 = A_1\,, && 2 (P \! \cdot \! x) {\cal A}_2 = - A_1 + A_2 -  A_3\,, 
\nonumber 
\\
2 {\cal A}_3 = A_3\,, &&  4 (P \! \cdot \! x) {\cal A}_4  = - 2 A_1 - A_3 - A_4  + 2 A_5\,, 
\nonumber 
\\
 4 (P \! \cdot \! x) {\cal A}_5 = A_3 - A_4\,, &&  4 \left(P \! \cdot \! x\right)^2  {\cal A}_6  =  A_1 - A_2 +  A_3 +  A_4 - A_5 + A_6\,,
\nonumber 
\\
{\cal T}_1 = T_1\,,  && 2 (P \! \cdot \! x) {\cal T}_2 = T_1 + T_2 - 2 T_3\,, 
\nonumber 
\\
2 {\cal T}_3 = T_7\,, &&  2 (P \! \cdot \! x) {\cal T}_4  = T_1 - T_2 - 2  T_7\,,
\nonumber 
\\
2 (P \! \cdot \! x) {\cal T}_5 = - T_1 + T_5 + 2  T_8\,, &&
4 \left(P \! \cdot \! x\right)^2 {\cal T}_6  = 2 T_2 - 2 T_3 - 2 T_4 + 2 T_5 + 2 T_7 + 2 T_8\,,
\nonumber 
\\
4 (P \! \cdot \! x) {\cal T}_7 =  T_7 - T_8\,, && 
4 \left(P \! \cdot \! x\right)^2 {\cal T}_8 = -T_1 + T_2 + T_5 - T_6 + 2 T_7 + 2 T_8 \,.
\nonumber
\\
\end{eqnarray}
The calligraphic notation is used for distribution amplitudes
belonging to a simple Dirac structure, while the non-calligraphic functions
denote distribution amplitudes of definite twist.
Each distribution amplitude $F = V_i,A_i,T_i,S_i,P_i$ can be represented  as
\begin{equation}
F(a_1, a_2, a_3, (P \! \cdot \! x)) = \int \! {\cal D} x\, e^{-i (P  \cdot  x) \sum_i x_i a_i} F(x_i)\,,
\end{equation}
where the functions $F(x_i)$ depend on the dimensionless
variables $x_i,\, 0 < x_i < 1, \sum_i x_i = 1$ which
correspond to the longitudinal momentum fractions
carried by the quarks inside the nucleon.
\section{Expansion of the nucleon distribution amplitudes up to next-to leading conformal spin}
In \cite{BFMS2000} the distribution amplitudes were expanded up to next-to leading order in the conformal spin.
The twist-3 distribution amplitudes read
\begin{eqnarray}
\label{twist-3}
V_1(x_i,\mu) &=& 120 x_1 x_2 x_3 \left[\phi_3^0(\mu) +
\phi_3^+(\mu) (1- 3 x_3)\right]\,,
\nonumber \\
A_1(x_i,\mu) &=& 120 x_1 x_2 x_3 (x_2 - x_1) \phi_3^-(\mu) \,,
\nonumber \\
T_1(x_i,\mu) &=& 120 x_1 x_2 x_3
\Big[\phi_3^0(\mu) - \frac12\left(\phi_3^+ - \phi_3^-\right)(\mu)
(1- 3 x_3)\Big]\,.
\end{eqnarray}
The twist-4 distribution amplitudes read
\begin{eqnarray}
\label{twist-4}
V_2(x_i,\mu)  &=& 24 x_1 x_2 \left[\phi_4^0(\mu)  + \phi_4^+(\mu)  (1- 5 x_3)\right] \,,
\nonumber\\
A_2(x_i,\mu)  &=& 24 x_1 x_2 (x_2 - x_1) \phi_4^-(\mu) \,,
\nonumber \\
T_2(x_i,\mu) &=& 24 x_1 x_2 \left[ \xi_4^0(\mu) + \xi_4^+(\mu)( 1- 5 x_3)\right]\,,
\nonumber \\
V_3(x_i,\mu)  &=&  12 x_3 \left[ \psi_4^0(\mu)(1-x_3)  + \psi_4^+(\mu)( 1-x_3 - 10 x_1 x_2) \right.
\nonumber\\
&& \left.\hspace*{+1.2cm}+ \psi_4^-(\mu) (x_1^2 + x_2^2 - x_3 (1-x_3) ) \right]\,,
\nonumber \\
A_3(x_i,\mu) &=& 12 x_3 (x_2-x_1) \left[\left(\psi_4^0 + \psi_4^+ \right)(\mu) + \psi_4^-(\mu) (1-2 x_3) \right] \,,
\nonumber \\
T_3(x_i,\mu)  &=&
6 x_3 \left[
(\phi_4^0 + \psi_4^0 + \xi_4^0)(\mu)(1-x_3) \right.\nonumber\\
&&\hspace*{+1.2cm}\left.+
(\phi_4^+ + \psi_4^+ + \xi_4^+)(\mu) ( 1-x_3 - 10 x_1 x_2)
\right. \nonumber \\
&&\left. \hspace*{+1.2cm}+
(\phi_4^- - \psi_4^- + \xi_4^-)(\mu) (x_1^2 + x_2^2 - x_3 (1-x_3) )
\right],
\nonumber \\
T_7(x_i,\mu)  &=&
6 x_3 \left[
(\phi_4^0 + \psi_4^0 - \xi_4^0)(\mu)(1-x_3) \right.\nonumber\\
&&\left.\hspace*{+1.2cm}+
(\phi_4^+ + \psi_4^+ - \xi_4^+)(\mu) ( 1-x_3 - 10 x_1 x_2)
\right. \nonumber \\
&&\left.\hspace*{+1.2cm} +
(\phi_4^- - \psi_4^- - \xi_4^-)(\mu) (x_1^2 + x_2^2 - x_3 (1-x_3) )
\right],
\nonumber \\
S_1(x_i,\mu) &=&
6 x_3 (x_2-x_1) \left[
(\phi_4^0 + \psi_4^0 + \xi_4^0 + \phi_4^+ + \psi_4^+ + \xi_4^+)(\mu)
\right. \nonumber\\
&& \left.\hspace*{+2.2cm} + (\phi_4^- - \psi_4^- + \xi_4^-)(\mu)(1-2 x_3) \right],
\nonumber \\
P_1(x_i,\mu) &=&
6 x_3 (x_1-x_2) \left[
(\phi_4^0 + \psi_4^0 -\xi_4^0 + \phi_4^+ + \psi_4^+ - \xi_4^+)(\mu)
\right. \nonumber\\
&& \left.\hspace*{+2.2cm} + (\phi_4^- - \psi_4^- - \xi_4^-)(\mu)(1-2 x_3) \right].
\end{eqnarray}
The twist-5 amplitudes are given by
\begin{eqnarray}
\label{twist-5}
V_4(x_i,\mu) &=& 3 \left[
\psi_5^0(\mu)(1-x_3) + \psi_5^+(\mu)(1-x_3 - 2 (x_1^2 +  x_2^2))
\right.
\nonumber\\&&\left.\hspace*{+1.2cm}{}
+ \psi_5^-(\mu)\left(2 x_1x_2 - x_3(1-x_3)\right) \right],
\nonumber\\
A_4(x_i,\mu) &=& 3 (x_2 -x_1) \left[- \psi_5^0(\mu) + \psi_5^+(\mu)(1- 2 x_3)
+ \psi_5^-(\mu) x_3  \right],
\nonumber \\
T_4(x_i,\mu) &=& \frac32 \left[
(\phi_5^0 +  \psi_5^0 + \xi_5^0) (\mu)(1-x_3)
\right.\nonumber\\&&\left.\hspace*{1.2cm}{}
+ \left(\phi_5^+ + \psi_5^+ + \xi_5^+ \right)(\mu)(1-x_3 - 2 (x_1^2 +  x_2^2))
\right.\nonumber\\&&\left.\hspace*{1.2cm}{}
+\left(\phi_5^- - \psi_5^- + \xi_5^- \right) (\mu)\left(2 x_1x_2 - x_3(1-x_3)\right)\right],
\nonumber \\
T_8(x_i,\mu) &=& \frac32 \left[
(\phi_5^0 +  \psi_5^0 - \xi_5^0) (\mu)(1-x_3)
\right.\nonumber\\&&\left.\hspace*{1.2cm}{}
+ \left(\phi_5^+ + \psi_5^+ - \xi_5^+ \right)(\mu)(1-x_3 - 2 (x_1^2 +  x_2^2))
\right.\nonumber\\&&\left.\hspace*{1.2cm}{}
+\left(\phi_5^- - \psi_5^- - \xi_5^- \right) (\mu)\left(2 x_1x_2 - x_3(1-x_3)\right)
\right],
\nonumber \\
V_5(x_i,\mu) &=& 6 x_3
\left[\phi_5^0(\mu)  + \phi_5^+(\mu)(1- 2 x_3)\right],
\nonumber\\
A_5(x_i,\mu) &=& 6 x_3 (x_2-x_1) \phi_5^-(\mu) \,,
\nonumber\\
T_5(x_i,\mu) &=& 6 x_3 \left[
\xi_5^0(\mu) + \xi_5^+(\mu)( 1- 2 x_3)\right],
\nonumber \\
S_2(x_i,\mu) &=& \frac32 (x_2 -x_1)
\left[- \left(\phi_5^0 + \psi_5^0 + \xi_5^0\right)(\mu)
\right.\nonumber\\&&\left.\hspace*{1.2cm}{}
+ \left(\phi_5^+ + \psi_5^+ + \xi_5^+\right)(\mu) (1- 2 x_3)
\right.\nonumber\\&&\left.\hspace*{1.2cm}{}
+ \left(\phi_5^- - \psi_5^- + \xi_5^- \right)(\mu) x_3
\right],
\nonumber \\
P_2(x_i,\mu) &=& \frac32 (x_2 -x_1)
\left[- \left(-\phi_5^0 - \psi_5^0 + \xi_5^0\right)(\mu)
\right.\nonumber\\&&\left.\hspace*{1.2cm}{}
+ \left(-\phi_5^+ - \psi_5^+ + \xi_5^+\right)(\mu) (1- 2 x_3)
\right.\nonumber\\&&\left.\hspace*{1.2cm}{}
+ \left(-\phi_5^- + \psi_5^- + \xi_5^- \right)(\mu) x_3
\right]\,,
\end{eqnarray}
and the twist-6 contributions are given by
\begin{eqnarray}
\label{twist-6}
V_6(x_i,\mu) &=& 2 \left[\phi_6^0(\mu) +  \phi_6^+(\mu) (1- 3 x_3)\right],
\nonumber \\
A_6(x_i,\mu) &=& 2 (x_2 - x_1) \phi_6^- \,, \\
T_6(x_i,\mu) &=& 2 \Big[\phi_6^0(\mu) -
\frac12\left(\phi_6^+-\phi_6^-\right) (1\!-\! 3 x_3)\Big].
\nonumber
\end{eqnarray}
The coefficients $\phi_i^x, \psi_i^x$ and $\xi_i^x$ ($i$ stands for the twist) in the above expansions 
can be expressed in terms of the eight
non-perturbative parameters $f_N, \lambda_1, \lambda_2, f_1^u, f_1^d, f_2^d, A_1^u, V_1^d$, defined in
section 2.
The corresponding relations read, for the leading conformal spin:
\begin{eqnarray}
\phi_3^0 = \phi_6^0 = f_N \,,\hspace{0.15cm}
&~&
\phi_4^0 = \phi_5^0 =
\frac{1}{2} \left(f_N + \lambda_1 \right) \,,
\nonumber \\
\xi_4^0 = \xi_5^0 =
\frac{1}{6} \lambda_2\,,
&~&
\psi_4^0  = \psi_5^0 =
\frac12\left(f_N - \lambda_1 \right).
\end{eqnarray}
For the next-to-leading spin,
for twist-3:
\begin{eqnarray}
\phi_3^- = \frac{21}{2} f_N \, A_1^u, &~&
\phi_3^+ = \frac{7}{2}  f_N \, (1 - 3 V_1^d),
\end{eqnarray}
for twist-4:
\begin{eqnarray}
\phi_4^+ &=& \frac{1}{4} \left[
 f_N( 3 - 10  V_1^d ) + \lambda_1(3- 10 f_1^d)
\right],
\nonumber \\
\phi_4^- &=& - \frac{5}{4} \left[
f_N( 1 - 2 A_1^u ) - \lambda_1(1- 2 f_1^d -4 f_1^u)
\right],
\nonumber \\
\psi_4^+ &=& - \frac{1}{4} \left[
f_N( 2\! +\! 5 A_1^u \!-\! 5 V_1^d) - \lambda_1 (2 \!-\! 5 f_1^d \!-\! 5 f_1^u)
\right],
\nonumber \\
\psi_4^- &=&  \frac{5}{4} \left[
f_N(2 - A_1^u - 3 V_1^d ) - \lambda_1(2- 7 f_1^d + f_1^u)
\right],
\nonumber \\
\xi_4^+ &=& \frac{1}{16} \lambda_2 (4\!-\! 15 f_2^d)\,, ~
\xi_4^- = \frac{5}{16} \lambda_2(4\!-\! 15 f_2^d),
\end{eqnarray}
for twist-5:
\begin{eqnarray}
\phi_5^+ &=&
- \frac{5}{6} \left[
f_N( 3 + 4   V_1^d) - \lambda_1 (1 - 4 f_1^d )
\right],
\nonumber \\
\phi_5^- &=&
- \frac{5}{3} \left[
f_N( 1 - 2 A_1^u ) - \lambda_1(f_1^d - f_1^u)
\right],
\nonumber \\
\psi_5^+ &=&
-\frac{5}{6} \left[
f_N(5\! + \!2 A_1^u \!-\!2 V_1^d) - \lambda_1 (1 \!-\! 2 f_1^d \!-\! 2 f_1^u)
\right],
\nonumber \\
\psi_5^- &=& \phantom{-}
\frac{5}{3} \left[
f_N( 2 - A_1^u - 3 V_1^d) + \lambda_1 (f_1^d - f_1^u)
\right],
\nonumber \\
\xi_5^+ &=& \phantom{-}\frac{5}{36} \lambda_2 (2- 9 f_2^d) \,,
\; \; \; \;
\xi_5^- = - \frac{5}{4} \lambda_2 f_2^d\,,
\end{eqnarray}
and for twist-6:
\begin{eqnarray}
\phi_6^+ &=& \frac{1}{2}\left[
f_N ( 1 - 4 V_1^d ) - \lambda_1  (1 - 2 f_1^d)
\right],
 \\
\phi_6^- &=& \phantom{-}\frac{1}{2} \left[
f_N (1 +  4 A_1^u ) + \lambda_1 (1- 4 f_1^d - 2 f_1^u)
\right] \,.
\nonumber
\end{eqnarray}
\subsubsection*{$x^2$-corrections:}
Next we summarize the expressions for the $x^2$-corrections to the leading twist 
distribution amplitudes $V_1$, $A_1$ and $T_1$. These corrections have been determined in
\cite{BLMS2001,HW2004,BLW2006,L2007c}.
\\
For $V_1$ we have
\begin{eqnarray}
{\cal V}_1^{M(u)} (x_2) &=& \int \limits_0^{1-x_2} dx_1 V_1^{M}(x_1, x_2,1-x_1-x_2)
=
 \frac{x_2^2}{24} \left( f_N C_{f}^u + \lambda_1 C_{\lambda}^u  \right)\,,
\nonumber
\\
 {\cal V}_1^{M(d)} (x_3) &=& \int \limits_0^{1-x_3} dx_1 V_1^{M}(x_1,1-x_1-x_3,x_3)
=  
\frac{x_3^2}{24} \left( f_N C_{f}^d + \lambda_1 C_{\lambda}^d  \right)
\nonumber
\\
\end{eqnarray}
with
\begin{eqnarray}
C_{f}^u       & = & (1 - x_2)^3 \big[113 + 495x_2 - 552x_2^2 - 10A_1^u(1 - 3x_2)
+  2V_1^d(113 - 951x_2 + 828x_2^2) \big],
\nonumber
\\
C_{\lambda}^u & = & - (1 \!-\! x_2)^3
              \big[13 - 20f_1^d + 3x_2 + 10f_1^u(1 \!-\! 3x_2)\big],
\nonumber
\\
C_{f}^d       & = & - (1 \!-\! x_3) \big[1441 + 505x_3 - 3371x_3^2 + 3405x_3^3 - 1104x_3^4  
\nonumber\\&&{}- 24V_1^d\big(207 \!-\! 3x_3 \!-\! 368x_3^2 \!+\! 412x_3^3
- 138x_3^4\big) \big] - 12(73 - 220V_1^d) \ln(x_3),
\nonumber
\\
C_{\lambda}^d & = &  - (1-x_3) \Big[11 + 131x_3 - 169x_3^2 + 63x_3^3
-  30f_1^d (3 + 11x_3 - 17x_3^2 + 7x_3^3) \big]
\nonumber
\\
& &
- 12(3 - 10f_1^d) \ln(x_3)\,.
\end{eqnarray}
In the case of $A_1$ one finds
\begin{eqnarray}
{\cal A}_1^{M(u)} (x_2) &=& \int \limits_0^{1-x_2} dx_1 A_1^{M}(x_1, x_2,1-x_1-x_2)
      =  \frac{ x_2^2}{24} (1 - x_2)^3 \left( f_N D_{f}^u  + \lambda_1 D_{\lambda}^u  \right)\,,
\nonumber \\
{\cal A}_1^{M(d)} (x_3) &=& \int \limits_0^{1-x_3} dx_1 A_1^{M}(x_1,1-x_1-x_3,x_3) = 0
\,,
\end{eqnarray}
with
\begin{eqnarray}
D_{f}^u       & = & 11 + 45 x_2 - 2 A_1^u (113 - 951 x_2 + 828 x_2^2 ) 
+ 10 V_1^d (1 - 30 x_2)\,,
\nonumber\\
D_{\lambda}^u & = &  29 - 45 x_2 - 10 f_1^u (7 - 9 x_2) - 20 f_1^d (5 - 6 x_2)\,.
\nonumber\\
\end{eqnarray}
Finally, for $T_1$ one has
\begin{eqnarray}
{\cal T}_1^{M(u)} (x_2)  & = & \int \limits_0^{1-x_2} dx_1 T_1^{M}(x_1, x_2,1-x_1-x_2) = 
\frac{ x_2^2}{48}  \left( f_N E_{f}^u  + \lambda_1 E_{\lambda}^u  \right),
\nonumber 
\\
{\cal T}_1^{M(d)} (x_3)  & = & \int \limits_0^{1-x_3} dx_1 T_1^{M}(x_1,1-x_1-x_3,x_3) = 
\frac{ x_3^2 (1\!-\!x_3)^4}{4}  \left( f_N E_{f}^d  + \lambda_1 E_{\lambda}^d  \right)
\nonumber 
\\
\end{eqnarray}
with
\begin{eqnarray}
E_{f}^u & = & -\Big[
(1 - x_2)
\left(
3 (439 + 71 x_2 - 621 x_2^2 + 587 x_2^3 - 184 x_2^4)
\right.
\nonumber\\&&{}
+ 4 A^u_1 (1 - x_2)^2 (59 - 483 x_2 + 414 x_2^2)
\left.
- 4 V^d_1 (1301 - 619 x_2 - 769 x_2^2 + 1161 x_2^3 - 414 x_2^4)
\right)
\Big]
\nonumber\\&&{}
- 12(73 - 220V_1^d) \ln(x_2)\,,
\nonumber \\
E_{\lambda}^u & = & -\Big[
(1 - x_2)
( 5 - 211 x_2 + 281 x_2^2 - 111 x_2^3
\nonumber\\&&{} +10 (1 +  61 x_2 -  83 x_2^2 +  33 x_2^3) f_1^d - 40(1 - x_2)^2 (2 - 3 x_2) f_1^u)
\Big]
- 12 (3 - 10 f_1^d) \ln(x_2)\,,
\nonumber \\
E_{f}^d & = & 17 + 92 x_3 + 12 (A^u_1+V^d_1) (3 - 23 x_3)
\,,
\nonumber \\
E_{\lambda}^d & = & -7 + 20 f^d_1 + 10 f^u_1
\,.
\end{eqnarray}

\section{Expansion of the nucleon distribution amplitude of twist-3 up to next-to-next-to leading conformal spin}


The expansion of the leading-twist distribution amplitude in a basis which is diagonal 
with respect to one-loop renormalization reads  up to next-to-next-to-leading conformal
spin
\begin{align}
\varphi (x_1,x_2,x_3,\mu) = 120 x_1 x_2 x_3 f_N (\mu_0) L^{\frac{2}{3\beta_0}}
 \Big\{  1    & + h_{10} (\mu_0) (x_1 \! - 2 x_2 +x_3) L^{\frac{8}{3\beta_0}} 
\nonumber
\\ 
&  
+ h_{11} (\mu_0) (x_1 \! - x_3) L^{\frac{20}{9\beta_0}}
\nonumber
\\ 
&+ h_{20} (\mu_0) 
\left[ 1 + 7 (x_2 - 2 x_1 x_3 - 2 x_2^2) \right] L^{\frac{14}{3\beta_0}}
\nonumber
\\ 
& + h_{21}  (\mu_0)
\left( 1 - 4 x_2 \right) \left( x_1 - x_3 \right) L^{\frac{40}{9\beta_0}}
\nonumber
\\ 
&  + h_{22}  (\mu_0)
\left[ 3 - 9 x_2 + 8 x_2^2 - 12 x_1 x_3 \right] L^{\frac{32}{9 \beta_0}}
\Big \} \, 
\label{tw3dwave}
\end{align}
with
\begin{equation}
 L=\frac{\alpha_s(\mu)}{\alpha_s(\mu_0)}\, , \hspace{1cm} \beta _0 = 11 - \frac23 n_F \, .
\end{equation} 
The coefficients $h_{ij}$ can be expressed in terms of the moments by
\begin{align}
h_{10} (\mu)& = \frac{7}{2} \left( 1 - 3 \varphi^{010}(\mu)  \right) 
\\
& = - \frac{7}{4} \left(1 -3  \left( A_1^u (\mu) + V_1^d (\mu)\right)\right)
\\
       & =- \frac12 \left( \tilde{\phi}_3^+ (\mu)- \tilde{\phi}_3^- (\mu)\right) \, ,
\\
h_{11} (\mu)& = \frac{21}{2} \left( \varphi^{100} (\mu)-  \varphi^{001} (\mu)\right) 
\\
& = \frac{21}{4} \left( 1 + A_1^u (\mu)- 3 V_1^d (\mu)\right) 
\\
& = \frac12 \left( 3 \tilde{\phi}_3^+ (\mu)+\tilde{\phi}_3^- (\mu)\right) \, ,
\\
h_{20} (\mu)& =  \frac{18}{5} \left(h_{10}(\mu)
                            +4-7\left(3\varphi^{101}(\mu)+\varphi^{200}(\mu)+\varphi^{002} (\mu)\right)
                            \right) \, ,
\\
h_{21} (\mu)& = 126  \left( \varphi^{200}(\mu)-\varphi^{002} (\mu)\right) 
                            -9 h_{11} (\mu) \, ,
\\
h_{22} (\mu)& = \frac{21}{5}  \left(
                            -h_{10}(\mu)-4
                            +6 \left(\varphi^{101} (\mu)+ 2\varphi^{200} (\mu)+ 2\varphi^{002}(\mu)\right)
                         \right) \, .
\end{align}
Of course, this form of $\varphi$ is not uniquely determined by the moments. 
The anomalous dimensions were obtained, e.g., in \cite{ano1,ano2,ano3}.
\\
One can also write down the renormalization group equations for the moments 
$\varphi^{n_1n_2n_3}=V^{n_1n_2n_3}_1-A^{n_1n_2n_3}_1$
alone:
\begin{align}
 \varphi^{100}(\mu)=&\frac{1}{21} \left( 7 +   h_{10}(\mu_0) L^{\frac{8}{3\beta_0}} + h_{11}(\mu_0) L^{\frac{20}{9\beta_0}}\right)
 \, ,
 \\
 \varphi^{010}(\mu)=&\frac{1}{21} \left( 7 - 2 h_{10}(\mu_0) L^{\frac{8}{3\beta_0}} \right)
 \, ,
 \\
 \varphi^{001}(\mu)=&\frac{1}{21} \left( 7 +   h_{10}(\mu_0) L^{\frac{8}{3\beta_0}} - h_{11}(\mu_0) L^{\frac{20}{9\beta_0}}\right)
 \, ,
 \\
 \varphi^{101}(\mu)=&\frac{1}{126}\left(12 + 3 h_{10} (\mu_0)L^{\frac{8}{3\beta_0}}  
                                      - 2 h_{20}(\mu_0) L^{\frac{14}{3\beta_0}}                                 -   h_{22}(\mu_0) L^{\frac{32}{9\beta_0}}
                             \right) \, ,
  \\
 \varphi^{200}(\mu)=&\frac{1}{252}   \left(36 + 9 h_{10}(\mu_0) L^{\frac{8}{3\beta_0}} + 9 h_{11} (\mu_0)L^{\frac{20}{9\beta_0}}
                             \right.
                             \nonumber\\
                      &\qquad\left.
                                      +   h_{20} (\mu_0)L^{\frac{14}{3\beta_0}} + h_{21} (\mu_0)L^{\frac{40}{9\beta_0}} + 3 h_{22} (\mu_0)L^{\frac{32}{9\beta_0}}
                             \right) \, ,
  \\
 \varphi^{002}(\mu)=&\frac{1}{252}  \left(36  + 9 h_{10}(\mu_0) L^{\frac{8}{3\beta_0}} - 9 h_{11} (\mu_0)L^{\frac{20}{9\beta_0}}
                             \right.
                             \nonumber\\
                      &\qquad\left.
                                      +   h_{20}(\mu_0) L^{\frac{14}{3\beta_0}} - h_{21} (\mu_0)L^{\frac{40}{9\beta_0}} + 3 h_{22}(\mu_0) L^{\frac{32}{9\beta_0}}
                             \right) \, .
\end{align}
Next we determine $V_1$, $A_1$ and $T_1$ from $\varphi$ up to $d$-wave contributions.
Including all anomalous dimensions we obtain (in the following we suppress the explicit renormalization 
scale dependence in the formulas)
\begin{align}
V_1 (x_1,x_2,x_3) = 120 x_1 x_2 x_3 f_N  L^{\frac{2}{3\beta_0}}
& \left\{ 1   - \frac{h_{10}}{2} (1 - 3 x_3) L^{\frac{8}{3\beta_0}} \right. 
  + \frac{h_{11}}{2} (1- 3 x_3) L^{\frac{20}{9\beta_0}}
\nonumber
\\ 
& - \frac{h_{20}}{2} 
\left[ (-2 + 7 (x_1 + x_2 - 4 x_1 x_2) \right] L^{\frac{14}{3\beta_0}}
\nonumber
\\ 
& -\frac{h_{21}}{2} 
\left[ -1 + 8x_2 - 8 x_2^2 - x_3 - 8 x_2 x_3 + 4 x_3^2 \right]
L^{\frac{40}{9\beta_0}}
\nonumber
\\ 
& \left. + \frac{h_{22}}{2} 
\left[ 6 - 21 x_1 + 20 x_1^2 - 21x_2 + 24 x_1 x_2 + 20 x_2^2 \right] L^{\frac{32}{9 \beta_0}}
\right\} \, ,
\end{align}
\begin{align}
A_1 (x_1,x_2,x_3) = - 60 x_1 x_2 x_3 (x_1-x_2) f_N  L^{\frac{2}{3\beta_0}}
& \left\{ 3  h_{10} L^{\frac{8}{3\beta_0}} \right. 
  + h_{11} L^{\frac{20}{9\beta_0}}
\nonumber
\\ 
& \left. 
  + (1-4 x_3) \left( 7 h_{20} L^{\frac{14}{3\beta_0}}
                     + h_{21} L^{\frac{40}{9\beta_0}}
                     + h_{22} L^{\frac{32}{9\beta_0}} \right)
\right\} \, ,
\end{align}
\begin{align}
T_1 (x_1,x_2,x_3) = 120 x_1 x_2 x_3 f_N  L^{\frac{2}{3\beta_0}}
& \left\{ 1   +  h_{10} (1 - 3 x_3) L^{\frac{8}{3\beta_0}} \right. 
\nonumber
\\ 
& - h_{20} 
\left[ -1 + 14 x_1 x_2 - 7 x_3 + 14 x_3^2 \right] L^{\frac{14}{3\beta_0}}
\nonumber
\\ 
& \left. -  h_{22} 
\left[ -3 + 12 x_1 x_2 + 9 x_3 - 8 x_3^2 \right] L^{\frac{32}{9 \beta_0}}
\right\} \, .
\end{align}
In the light-cone sum rule determination of the nucleon form factors we need
the distribution amplitudes at a certain renormalization scale $\mu= \mu_0$, therefore we give also the
simplified  expressions ($L \equiv 1$) in the following.
\\
Our expressions agree up to next-to-leading conformal spin with the correspondings ones of
\cite{BFMS2000}. 
In \cite{BLMS2001} the vector function $V_1$ including the next-to-next-to-leading conformal spin was used with
the following notation:
\begin{align}
V_1 (x_1,x_2,x_3) &= 120 x_1 x_2 x_3 f_N 
\bigg[1 + \tilde{\phi}_3^+(\mu)(1-3x_3) +\tilde\phi_3^{d1}\big[3-21 x_3+28 x_3^2\big]
\nonumber \\
& \hspace{3cm}    
+\tilde\phi_3^{d2}\big[5(x_1^2+x_2^2)-3(1-x_3)^2)\big]\bigg]
\,,
\\
A_1 (x_1,x_2,x_3) & = 120 x_1 x_2 x_3 (x_2-x_1) f_N  
\left\{ \tilde{\phi}_3^-  + (1-4 x_3) \tilde{\phi}_3^{d3} \right\} \, ,
\\
T_1 (x_1,x_2,x_3,\mu) & = 120 x_1 x_2 x_3 f_N 
 \left\{ 1   +  \frac{1}{2} (\tilde{\phi}_3^- - \tilde{\phi}_3^+) (1 - 3 x_3) 
\right. 
\nonumber
\\ 
& \hspace{3cm} + \tilde{\phi}_3^{d4}
\left[ 1 - 14 x_1 x_2 + 7 x_3 - 14 x_3^2 \right] 
\nonumber
\\ 
& \hspace{3cm} \left. +  \tilde{\phi}_3^{d5}
\left[ 1 + x_1 x_2 - 8 x_3 + 11 x_3^2 \right] 
\right\} \, ,
\end{align}
with
\begin{align}
\tilde\phi_3^{d1} & = \frac{1}{10} \left(    h_{20} -   h_{21} + 3 h_{22} \right)
\\
&= \frac{9}{10}   \left(3 + 28 \varphi^{002} - 21 V^d_1 \right) \, ,
\\
\tilde\phi_3^{d2} &= \frac{1}{5}  \left(- 7 h_{20} + 2 h_{21} + 4 h_{22} \right)
\\
&= -\frac{63}{5}  \left( 3 + 5 A^u_1 - V^d_1 - 2 ( \varphi^{002} + 5  \varphi^{101} +5  \varphi^{200}) \right) \, ,
\\
\tilde{\phi}_3^{d3} & = \frac{1}{2} \left( 7 h_{20} + h_{21} + h_{22} \right)
\\
&= \frac{63}{2}    \left(A^u_1 + 4 V^d_1  - 4 (  \varphi^{002} +2  \varphi^{101}) \right) \, ,
\\
\tilde{\phi}_3^{d4} & = h_{20} + h_{22} 
\\
&= -\frac{9}{20}  \left(3 + 7 (A^u_1 + V^d_1) - 56 (  \varphi^{002} - 2  \varphi^{101} +  \varphi^{200}) \right) \, ,
\\
\tilde{\phi}_3^{d5} & = 2 h_{22} 
\\
&= - \frac{63}{10}  \left(3 + 7 (A^u_1 + V^d_1) - 8 (2  \varphi^{002} +  \varphi^{101} + 2  \varphi^{200}) \right) \, .
\end{align}
Numerically one obtains for the COZ model  \cite{COZ1987}
\begin{eqnarray}
 \tilde\phi_3^{d1}(\mu = 1 {\rm GeV}) &=& 0.61\,,
\\
 \tilde\phi_3^{d2}(\mu = 1 {\rm GeV}) &=& 3.7\,.
\label{twist3par}
\end{eqnarray}
This agrees with the numbers quoted in \cite{BLMS2001}.
Using the lattice calculation we get
\begin{eqnarray}
 \tilde\phi_3^{d1}(\mu = 1 {\rm GeV}) &=& 0.51\,,
\\
 \tilde\phi_3^{d2}(\mu = 1 {\rm GeV}) &=& 0.71\,.
\end{eqnarray}
Here again the pure QCD sum rule calculation seems to overstimate the effects.

\section{Models for the leading-twist nucleon distribution amplitude}
In this section we present concrete models for the leading-twist nucleon distribution amplitude including 
next-to-next-to-leading conformal spin at the renormalization scale $1\,\mathrm{GeV}$.
At twist-3 one independent distribution amplitude $\varphi(x_1,x_2,x_3,\mu)$ arises, see, e.g., \cite{BFMS2000}:
\begin{equation}
\varphi(x_1,x_2,x_3,\mu) =\left( V_1 - A_1 \right) (x_1,x_2,x_3,\mu) \, . 
\end{equation}
In \cite{BFMS2000} this distribution amplitude was denoted by $ \Phi_3(x_1,x_2,x_3,\mu)$.
From $\varphi$ one easily gets $V_1$, $A_1$ and $T_1$, see, e.g., \cite{BFMS2000}: 
\begin{align}
T_1 (x_1,x_2,x_3) &= \frac12 \left[\varphi(x_1,x_3,x_2)+\varphi(x_2,x_3,x_1) \right] \, ,
\\
V_1 (x_1,x_2,x_3) &= \frac12 \left[\varphi(x_1,x_2,x_3)+\varphi(x_2,x_1,x_3) \right] \, ,
\\
A_1 (x_1,x_2,x_3) &= \frac12 \left[\varphi(x_2,x_1,x_3)-\varphi(x_1,x_2,x_3) \right] \, .
\end{align}
The asymptotic form - only the leading conformal spin contribution - of  $\varphi(x_1,x_2,x_3,\mu)$
reads
\begin{equation}
\varphi_{Asy} (x_1,x_2,x_3,\mu) = 120 x_1 x_2 x_3 \phi_3^0 (\mu),\quad (\phi_3^0\equiv f_N) \, .
\end{equation}
Including next-to leading conformal spin one gets \cite{BFMS2000}
\begin{equation}
\varphi (x_1,x_2,x_3,\mu) = \varphi_{Asy} (x_1,x_2,x_3,\mu) 
\left[ 1 + \tilde{\phi}_3^- (\mu) (x_1 - x_2) + \tilde{\phi}_3^+ (\mu) (1 - 3 x_3) \right] \, ,
\end{equation}
with
\begin{equation}
\tilde{\phi}_3^- = \frac{\phi_3^-}{\phi_3^0}
\,,
\hspace{2cm}
\tilde{\phi}_3^+ = \frac{\phi_3^+}{\phi_3^0}\,.
\end{equation}
In the literature also second moments of the leading-twist distribution amplitude
were determined with QCD sum rules \cite{Chernyak1984,King1987,COZ1987}.
With this information one can build models for $\varphi(x_1,x_2,x_3)$
at a certain renormalization scale $\mu$,
including next-to-next-to leading conformal spin.
We will use
the model from \cite{COZ1987} $\varphi^{COZ}(x_1,x_2,x_3)$
and         
the model from \cite{King1987} $\varphi^{KS}(x_1,x_2,x_3)$:
\begin{align}
\varphi^{COZ}(x_1,x_2,x_3)  & =  \varphi_{Asy} (x_1,x_2,x_3) 
\nonumber 
\\
&
\left[ 23.814 x_1^2 + 12.978 x_2^2 + 6.174 x_3^2 + 5.88 x_3 - 7.098
\right] \, ,
\\
\varphi^{KS}(x_1,x_2,x_3) & = \varphi_{Asy} (x_1,x_2,x_3) 
\nonumber 
\\
&
\left[ 20.16 x_1^2 + 15.12 x_2^2 + 22.68 x_3^2 - 6.72 x_3 + 1.68 (x_1 \! \! - \! x_2) \! - 5.04
\right] \, .
\end{align}
Bolz and Kroll derived a very simple model using some experimental constraints
\cite{Bolz1996}. Their model
for the leading-twist distribution amplitude reads
\begin{align}
\varphi^{BK}(x_1,x_2,x_3)  & =  \frac12 \varphi_{Asy} (x_1,x_2,x_3) (1+ 3 x_1) \, .
\end{align}
Based on the lattice calculations of $\varphi^{100}$, $\varphi^{001}$, 
$\varphi^{101}$, $\varphi^{200}$ and $\varphi^{002}$ in  
\cite{Lattice2007,Lattice2008} and Eq.~\eqref{tw3dwave} we have obtained the model
\begin{equation}
\begin{split}
 \varphi^{LAT}(x_1,x_2,x_3)=\varphi_{Asy} (x_1,x_2,x_3)  \big (&
  -0.401 + 29.214 x_1 - 44.542 x_2  + 7.664 x_3 \\
 &+ 12.561 x x_2 + 31.748 x_1 x_3 - 103.09 x_2 x_3\\
 &- 41.880 x_1^2 + 92.958 x_2^2 + 17.836 x_3^2
  \big ) \, .
\end{split}
\end{equation}

\end{document}